%% file: ms.tex
\documentclass[iop]{emulateapj}
\usepackage{graphicx}

\shorttitle{Protocluster Galaxies at $z=2.2$}
\shortauthors{Zirm et al.}

\def\spose#1{\hbox to 0pt{#1\hss}}
\def\simlt{\mathrel{\spose{\lower 3pt\hbox{$\mathchar"218$}}
     \raise 2.0pt\hbox{$\mathchar"13C$}}}
\def\simgt{\mathrel{\spose{\lower 3pt\hbox{$\mathchar"218$}}
     \raise 2.0pt\hbox{$\mathchar"13E$}}}

\begin{document}

\title{Internal Structure of Protocluster Galaxies: Accelerated
  Structural Evolution in Overdense Environments?\altaffilmark{1}}

\author{Andrew W. Zirm\altaffilmark{2}}

\author{Sune Toft\altaffilmark{2}}
%\email{sune@dark-cosmology.dk}

\author{Masayuki Tanaka\altaffilmark{3}}

\altaffiltext{1}{Based on observations with the NASA/ESA Hubble Space
Telescope, obtained at the Space Telescope Science Institute, which is
operated by the Association of Universities for Research in Astronomy,
Inc., under NASA contract NAS 5-26555}

\altaffiltext{2}{Dark Cosmology Centre, Niels Bohr Institute, University
of Copenhagen, Juliane Maries Vej 30, DK-2100 Copenhagen,
Denmark; {\tt azirm@dark-cosmology.dk; sune@dark-cosmology.dk}}

\altaffiltext{3}{Institute for the Physics and Mathematics of the
  Universe, The University of Tokyo, 5-1-5 Kashiwanoha, Kashiwa, Chiba
  277-8583, Japan;
  {\tt masayuki.tanaka@ipmu.jp}}

\begin{abstract}

  We present a high spatial-resolution {\it HST}/NICMOS imaging survey
  in the field of a known protocluster surrounding the powerful radio
  galaxy MRC1138-262 at $z=2.16$.  Previously, we have shown that this
  field exhibits a substantial surface overdensity of red $J-H$
  galaxies.  Here we focus on the stellar masses and galaxy effective
  radii in an effort to compare and contrast the properties of likely
  protocluster galaxies with their field counterparts and to look for
  correlations between galaxy structure and (projected) distance
  relative to the radio galaxy.

  We find a hint that quiescent, cluster galaxies are on average less
  dense than quiescent field galaxies of similar stellar mass and
  redshift.  In fact, we find only two (of nine) quiescent
  protocluster galaxies are of simliar density to the majority of the
  massive, quiescent compact galaxies (SEEDs) found in several field
  surveys.  Furthermore, there is some indication that the structural
  Sersic $n$ parameter is higher ($n \sim 3-4$) on average for cluster
  galaxies compared to the field SEEDs ($n \sim 1-2$) This result may
  imply that the accelerated galaxy evolution expected (and observed)
  in overdense regions also extends to structural evolution presuming
  that massive galaxies began as dense (low $n$) SEEDs and have
  already evolved to be more in line with local galaxies of the same
  stellar mass.
  
\end{abstract}


\section{Introduction}

The internal spatial and velocity distribution of stars is an
indicator of the manner in which galaxies have formed, assembled and
evolved.  In the local universe, tidal streams, shells and
kinematically distinct cores are examples of archeological clues to
past merging and formation events \citep*[e.g.,][]{Pengetal02,
  Emsellemetal07, vanDokkum05, BlantonMoustakas09}.  Even coarse
measures, such as the average stellar surface mass density within the
effective radius ($\Sigma_{50}$), correlate with the star formation
rate or the mean stellar age.  Giant elliptical galaxies have high
stellar mass per unit area (or volume) and show negligible current
star formation while more diffuse stellar disks and dwarf irregulars
are forming stars at sometimes prodigious rates per unit stellar mass
(specific star-formation rate; sSFR).
% This could be inferred from the observation that star forming
% galaxies generally have an exponential surface-brightness profile
% while early-types/spheroids follow the more sharply peaked
% $R^{1/4}$--law.
At higher redshift, analogous relations are already in place
\citep*[e.g.,][]{Franxetal08}.  While observationally it remains
difficult to separate high-redshift galaxies into classical
Hubble-types we can now photometrically determine redshifts, stellar
masses and galaxy sizes for large numbers of galaxies at $z \sim 2$.
Such studies \citep*[e.g.,][]{Zirmetal07, Toftetal07, vanDokkumetal08,
  Toftetal09, Williamsetal10, Moslehetal11} have found that quiescent
galaxies are in general more dense than their star-forming
counterparts.
% at low and intermediate total stellar masses.

The origin of this bi-modal distribution of galaxy properties is
unclear.  It is possible that the quiescent $z \sim 2$ galaxies are
more compact because they formed when the universe was smaller and
more dense and mergers were more gas-rich.  It also seems plausible
that the process which quenches star formation may be linked to a
morphological change.  Or perhaps the dominant formation processes
differ for galaxies with different present-day stellar masses
\citep*{KhochfarSilk06, KhochfarSilk09, KhochfarSilk11,
  DekelBirnboim08, Dekeletal09}

In general terms, dissipation should result in more compact stellar
cores than dissipationless assembly
\citep*[e.g.,][]{CiottiLanzoniVolonteri07, NaabJohanssonOstriker09,
  Oseretal10}.  Dense stellar systems, once formed, tend to persist
through successive (minor) mergers.  The relatively recent discovery
of quiescent, massive and compact galaxies at $z \sim 2$ (hereafter
Semi-Evolved Elephantine Dense galaxies or ``SEEDs'') implies that at
least some galaxies have their origin in high-redshift ($z \simgt 4$),
gas-rich mergers \citep*{Daddietal05, Zirmetal07, Toftetal07,
  vanDokkumetal08, Cimattietal08}.  These mergers resulted in many
stars being formed in a relatively small volume.  By $z \sim 2$ these
systems have low star-formation rates and relatively high stellar
masses in addition to their small sizes ($r_{e} \simlt 1$kpc).  The
SEEDs therefore have extreme internal stellar mass densities
%($\Sigma_{50}$, the average surface mass
%density within the half-light radius in $M_{\odot}$ ${\rm kpc}^{-2}$).
They do not fall on the size-mass relation defined by local galaxies.
Their evolution from $z\sim2$ to $z=0$ is therefore a puzzle.

The most massive galaxies in the present-day universe are located at
the centers of rich clusters.  These galaxy overdensities were
statistically the first to separate from the Hubble flow and collapse
and are therefore believed to follow an accelerated timeline for the
process of galaxy formation.  There is some observational evidence
that galaxies in the progenitors of clusters, protoclusters, do have
significantly older stars and higher masses than galaxies in the field
at similar redshifts \citep{Steideletal05, Tanakaetal10}.  Might
cluster galaxies, having formed earlier, be even more dense than field
SEEDs? Or, alternatively, the 'fast-forward' evolution of cluster
galaxies may lead to lower density galaxies in protoclusters compared
to their field counterparts.  It is interesting, then, to look for
dense SEED galaxies in protoclusters at redshift $z\sim2$.

We have undertaken a NICMOS imaging program to study the red galaxy
population in a protocluster at $z=2.16$.  Broad and narrow-band
imaging, both in the optical and near-infrared, of the field
surrounding the powerful radio galaxy MRC 1138-262 ($z=2.16$) have
identified more than 100 candidate companion galaxies.
% This target served as the proof-of-concept for the successful VLT
% Large Program summarized in Venemans et
% al. (2007)\nocite{Venemansetal07}.
There are surface-overdensities of both line-emitting candidates
(Lyman-$\alpha$ and H$\alpha$), X-ray point sources, sub-mm selected
galaxies and red optical--near-infrared galaxies
\citep*{Pentericcietal02,KurkPhD,Kurketal04,Stevensetal03,
  Croftetal05}.  Fifteen of the Ly$\alpha$ and 9 of the H$\alpha$
emitters have been spectroscopically confirmed to lie at the same
redshift as the radio galaxy \citep*{Kurketal04b}.
% The $I-K$-selected Extremely Red Objects (EROs; $I-K > 4.3$ Vega)
% seem concentrated around the RG but have no spectroscopic redshifts
% at this time.
By obtaining deep images through the NICMOS $J_{110}$ and
$H_{160}$ filters, which effectively span the 4000\AA-break at
$z=2.16$, we have identified a large surface overdensity of red
galaxies consistent with a forming red sequence \citep*{Zirmetal08}.
In this paper we present a more detailed analysis of the masses and
morphologies of galaxies in this field.  The article is organized as
follows: in Section \S\ref{sec:obs} we describe the data and their
reductions, in Section \S\ref{sec:analysis} we present the photometric
redshifts, stellar population models and morphological fits, in
Section \S\ref{sec:res} we present the internal stellar mass densities
and other derived properties and finally in \S\ref{sec:discussion} we
discuss these results in the context of galaxy evolution models.  We
use a $(\Omega_{\Lambda},\Omega_{M}) = (0.7,0.3)$, $H_{0} = 73$
${\rm km}$ ${\rm s^{-1}}$ ${\rm Mpc^{-1}}$ cosmology throughout.  At
$z=2.16$ one arcsecond is equivalent to 8.4 kpc.  All magnitudes are
referenced to the AB system \citep*{AB} unless otherwise noted.

\section{Observations and Data Reductions\label{sec:obs}}

\subsection{NICMOS Imaging\label{sec:NICMOS}}

The NICMOS instrument on-board {\it HST} is capable of deep
near-infrared imaging over a relatively small field-of-view
($51\arcsec \times 51\arcsec$).  In the case of MRC~1138-262, we know
that galaxies are overdense on the scale of a few arcminutes
\citep*{Kurketal04, Croftetal05} and are thus well-suited for
observations with NICMOS camera 3 on {\it HST}.  We used 30 {\it HST}
orbits to image seven overlapping pointings in both filters and one
additional pointing in $H_{160}$ alone.  These observations reach an
AB limiting magnitude ($m_{10\sigma}$; 10$\sigma$, $0\farcs5$ diameter
circular aperture) of $m_{10\sigma}=24.9$~mag in $J_{110}$ and
$m_{10\sigma}=25.1$~mag in $H_{160}$.  The same field was imaged in
the $g_{475}$ ($m_{10\sigma}=27.5$~mag) and $I_{814}$
($m_{10\sigma}=26.8$~mag) filters using the Wide-Field Channel of the
Advanced Camera for Surveys on {\it HST} as part of a Guaranteed Time
program (\# 10327; Miley et al. 2006\nocite{Mileyetal06}).

\begin{figure*}
\includegraphics[scale=0.75]{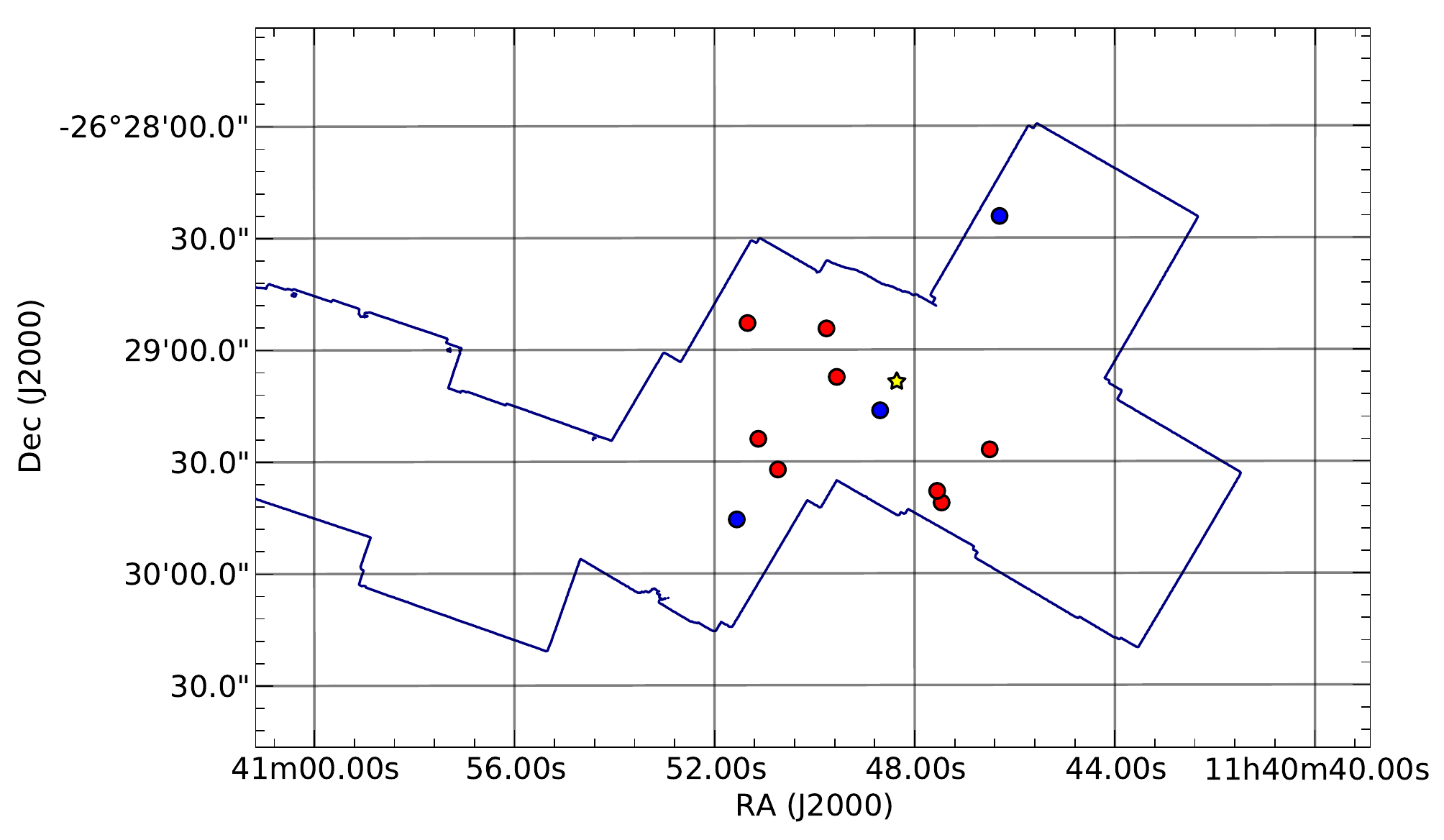}
\caption{Outline of the NICMOS mosaic. The red and blue points mark
  the locations of the quiescent and star-forming cluster galaxies
  respectively. The yellow star is the radio galaxy
  MRC1138-262.\label{fig:spatial}}
\end{figure*}

The NICMOS images were reduced using the on-the-fly reductions from
the {\it HST} archive, the IRAF task {\it pedsky} and the {\it
  dither/drizzle} package to combine the images in a mosaic.  The
dither offsets were calculated using image cross-correlation and were
refined iteratively.  Alignment of the pointings relative to each
other was accomplished using a rebinned version of the ACS $I_{814}$
image as a reference.  The final mosaic has a pixel scale of
$0\farcs1$.  Galaxies were selected using the $H_{160}$-band image for
detection within SExtractor \citep*{SExtractor}.  We used a
$2.2\sigma$ detection threshold with a minimum connected area of 10
pixels.  We also corrected the NICMOS data for the count-rate
dependent non-linearity \citep*{CPSNONLINEAR}.  Total galaxy
magnitudes were estimated by using the MAG\_AUTO values from
SExtractor. We show the outline of the NICMOS mosaic in Figure
\ref{fig:spatial} along with the positions of the radio galaxy (yellow
star) and star-forming (blue circles) and quiescent (red circles)
protocluster galaxies.

The $J_{110} - H_{160}$ colors were determined by running SExtractor
\citep*{SExtractor} in two-image mode using the $H_{160}$ image for
object detection and isophotal apertures.  The $J_{110}$ image was
PSF-matched to the $H_{160}$ band.  We also incorporated the two ACS
bands \citep*{Mileyetal06}, the Spitzer IRAC bands, $U_{n}$
\citep*{Zirmetal08} and $V$ bands from Keck/LRIS, $z$ and $R$ from
VLT/FORS2 \citep*{Kurketal04, Kurketal04b}, $H$ band from NTT/SOFI and
$J$ and $Ks$ from Subaru/MOIRCS \citep*{Kodamaetal07}.  The assembly
of the merged multi-band catalog is detailed in Tanaka et
al. 2010\nocite{Tanakaetal10}.

\subsection{FIREWORKS Survey Data and Literature Sample\label{sec:FIREWORKS}}

The FIREWORKS data are described in detail in Wuyts et al.\
(2008)\nocite{FIREWORKS}.  In brief, the survey is $K_{S}$-band
selected to $5\sigma$ depth of 24.3 (AB) over an area of
113$\square\arcmin$.  In addition to the deep $K_{S}$ band data there
is high-quality imaging in each of the $U$, $B$, $V$, $I$, $i$, $z$,
$J$, $H$, the four {\it Spitzer}/IRAC bands and the $24\mu$m {\it
  Spitzer}/MIPS band.  The combined multi-band catalog has been used
to measure precise photometric redshifts, galaxy sizes
\citep*{Toftetal09} and to model the spectral energy distributions to
derive stellar masses, ages and star-formation rates
\citep*{Damenetal09}.  We have made three cuts to the FIREWORKS sample
to ensure that we are making appropriate field-to-cluster
comparisons. First, since we are comparing galaxy sizes (densities) we
require that the galaxies are bright enough to have a reliable size
measurement in these ground-based data. Based on the comparison of
size measurements from VLT/ISAAC and HST/NICMOS for the same galaxies,
Toft et al. (2009) found that at $K\sim 21.5 -22.0$ the scatter
between these two size determinations increases significantly. We
therefore select only $K < 21.5$ galaxies from FIREWORKS. Next, we
have made a photometric redshift cut $1.9 < z < 2.6$ to select
galaxies within the field at roughly the same epoch as the
protocluster galaxies. Finally, we select the quiescent field
population on the basis of the specific star-formation rate (log(sSFR)
$< -11$ yr$^{-1}$). We note that after these cuts, the stellar
mass distribution remains similar to our protocluster galaxy
sample.

For further comparison to our protocluster field data, we have
compiled a sample of $z \sim 2$ quiescent galaxies published in the
literature. We used four references for this sample:
\citet*{Cassataetal10}, \citet*{vanDokkumetal08},
\citet*{Mancinietal10} and
\citet*{Saraccoetal09}. \citet*{Saraccoetal09} used {\it HST}/NICMOS
Camera 3 as we have, \citet*{Mancinietal10} used {\it HST}/ACS
imaging, \citet*{vanDokkumetal08} studied {\it HST}/NICMOS Camera 2
imaging while \citet*{Cassataetal10} use imaging from the WFC3/IR
channel on {\it HST}.

We have attempted to translate these published stellar mass estimates
to the same IMF (Salpeter) and to the same stellar population
synthesis model set (Maraston 2005). We have used the analyses of
Salimbeni et al. (2009; see their Fig.~1\nocite{SEDmodels}) to derive
mean corrections between model sets. The adopted IMF also affects the
derived star-formation rates. The offsets in this quantity are similar
in magnitude to the systematic shift in derived stellar mass
\citep*[e.g.,][]{Erbetal06, Nordonetal10}, so the specific star
formation rate (i.e., the ratio of star-formation rate to stellar
mass) should be effectively unchanged.

\begin{figure*}
\includegraphics[scale=0.4]{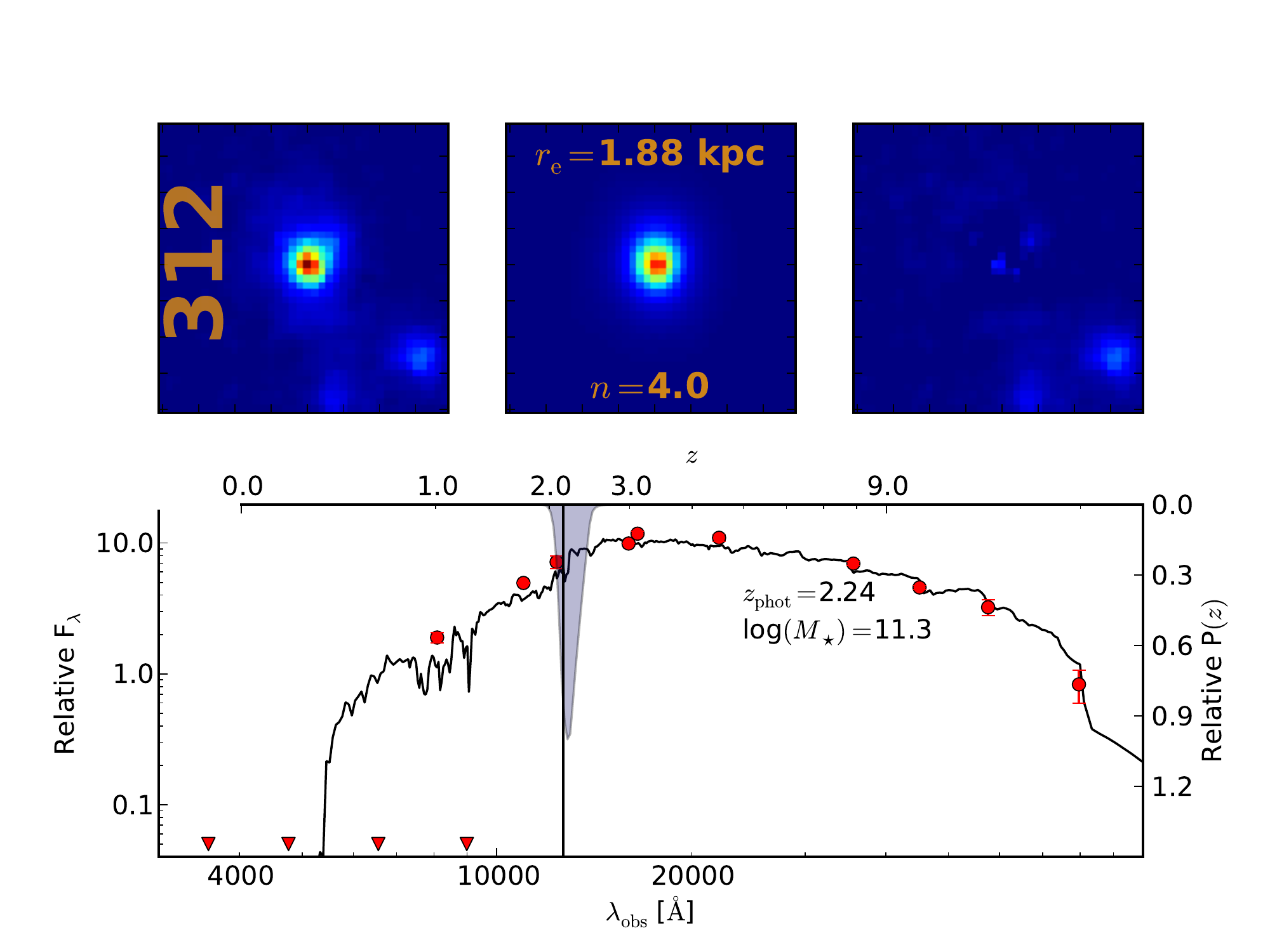}
\includegraphics[scale=0.4]{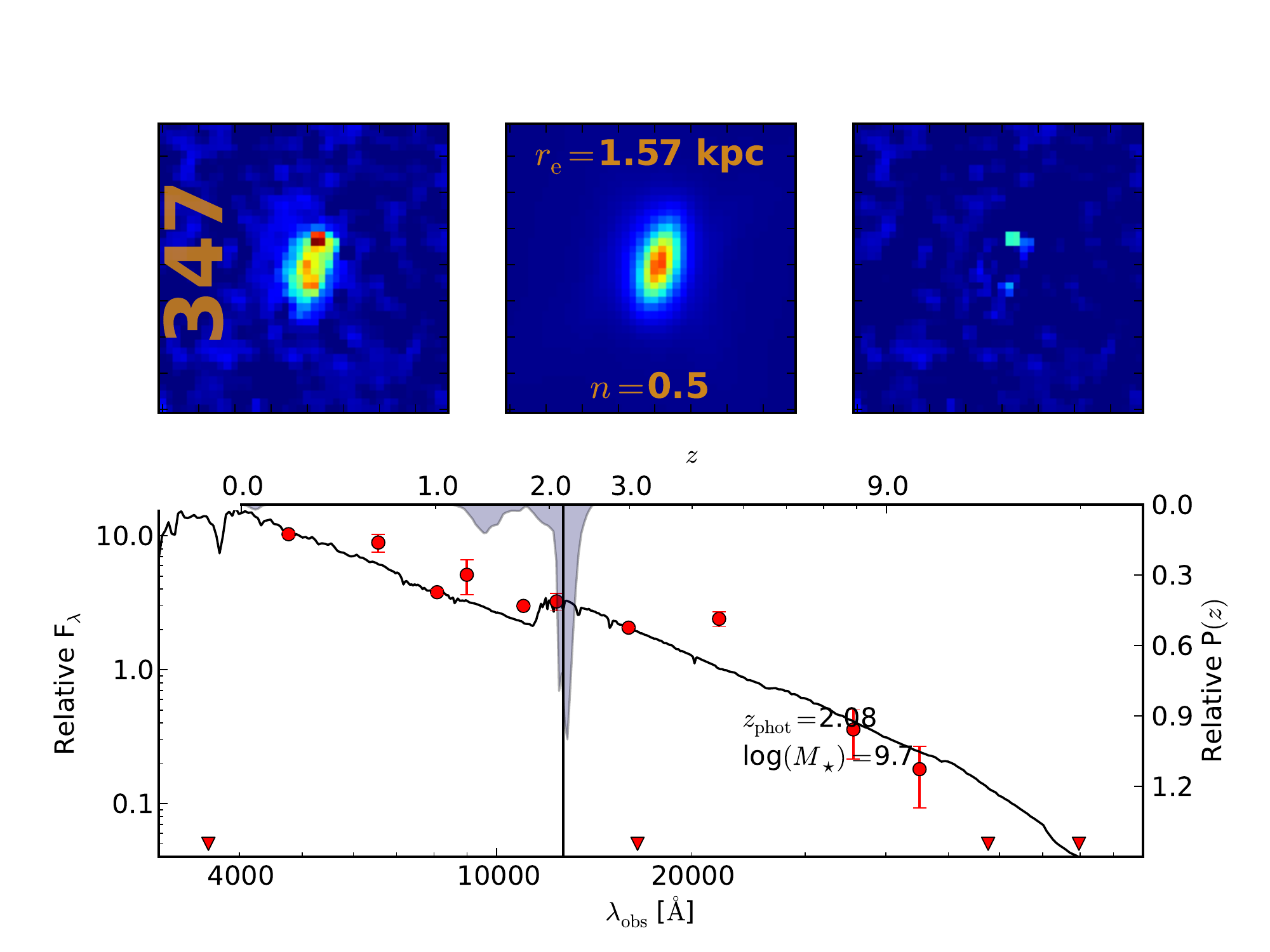}
\includegraphics[scale=0.4]{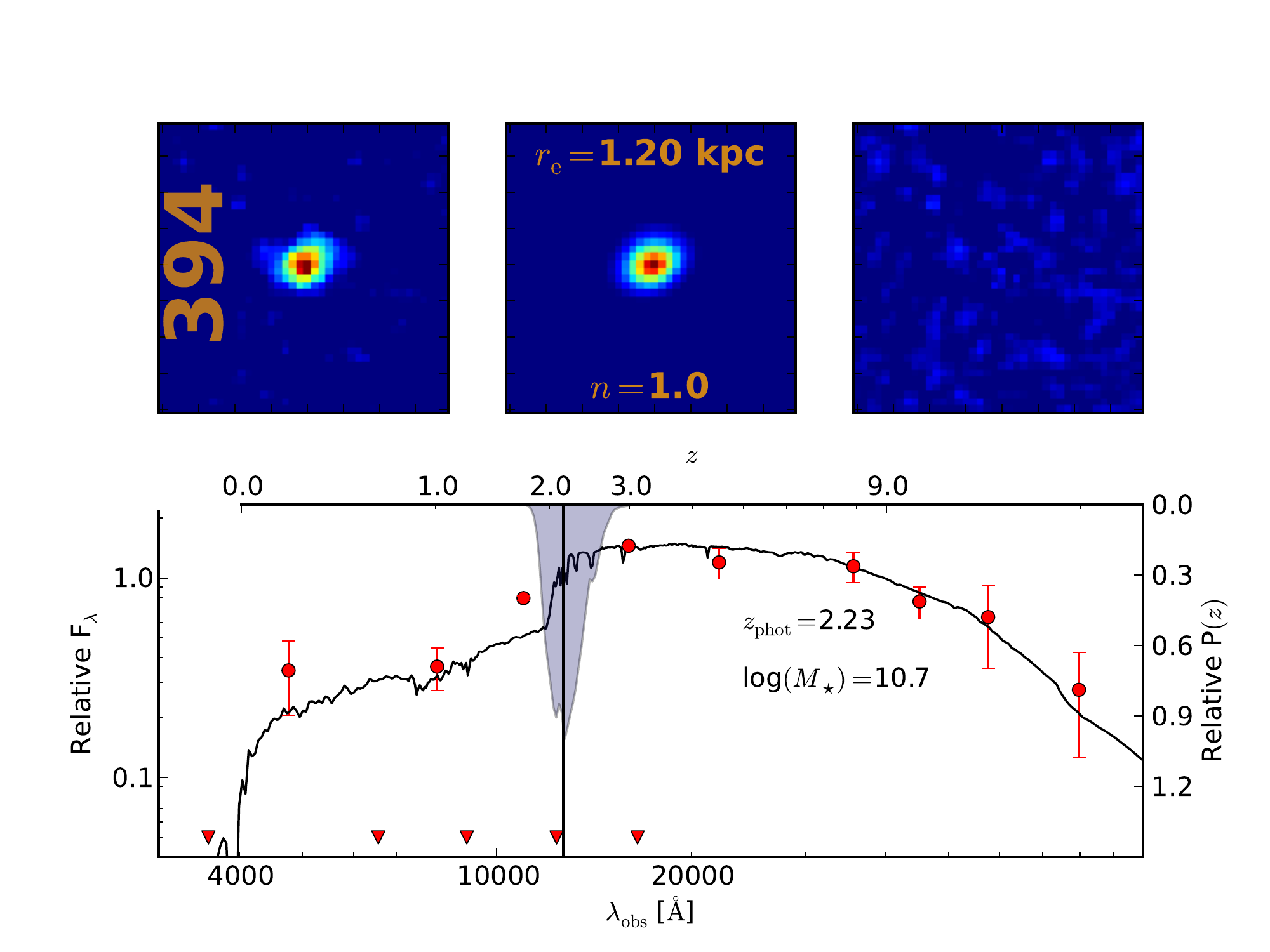}
\includegraphics[scale=0.4]{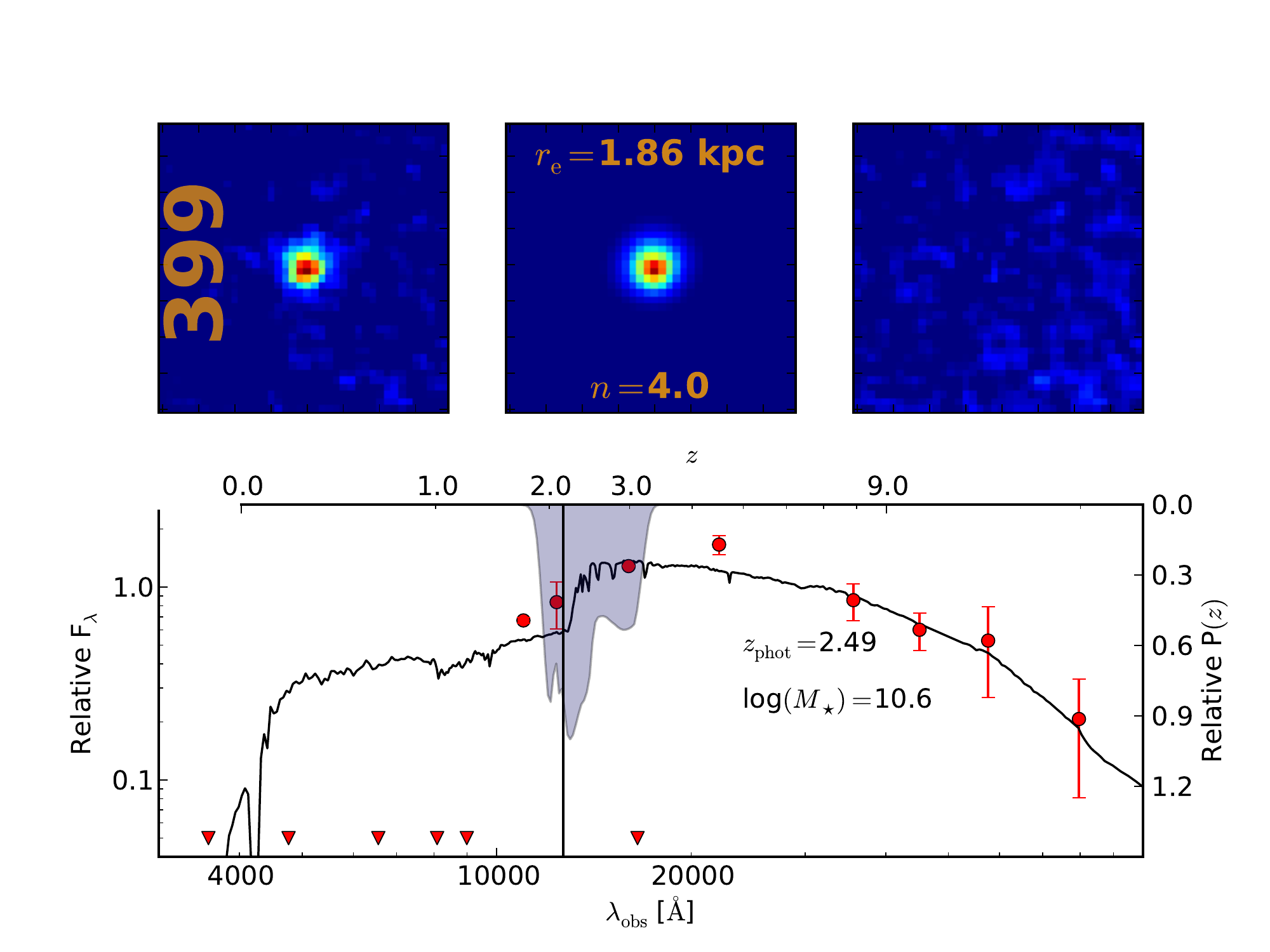}
\caption{Upper panels, left to right: Galaxy image, GALFIT best-fit
  model and image residuals after subtraction of the model.  Lower
  panel: Photometry and best-fit SED model from FAST \citep*[red
  circles and solid black line][]{FAST}.  The shaded regions represent
  the photometric redshift probability distribution (upper scale)
  centered at rest-frame 4000\AA.\label{fig:cuts1}}
\end{figure*}

\section{Analysis\label{sec:analysis}}

Here we combine the multiband photometric catalog and the NICMOS high
spatial-resolution imaging to derive physical parameters for
individual galaxies.  We pare down the total NICMOS galaxy sample to
those which have a high quality-of-fit for the photometric redshift
(using EAZY, \citet*{EAZY}), the spectral energy distribution (using
FAST, \citet{FAST}) and 2D surface-brightness profile fit (using
GALFIT, \citet*{GALFIT}).  This reduces the galaxy sample from the
$H_{160}$-band detected total of 711.  We further restrict our
attention to those galaxies which most likely lie within the known
protocluster (see Section \S\ref{sec:res:selection}).

\begin{figure*}
\includegraphics[scale=0.4]{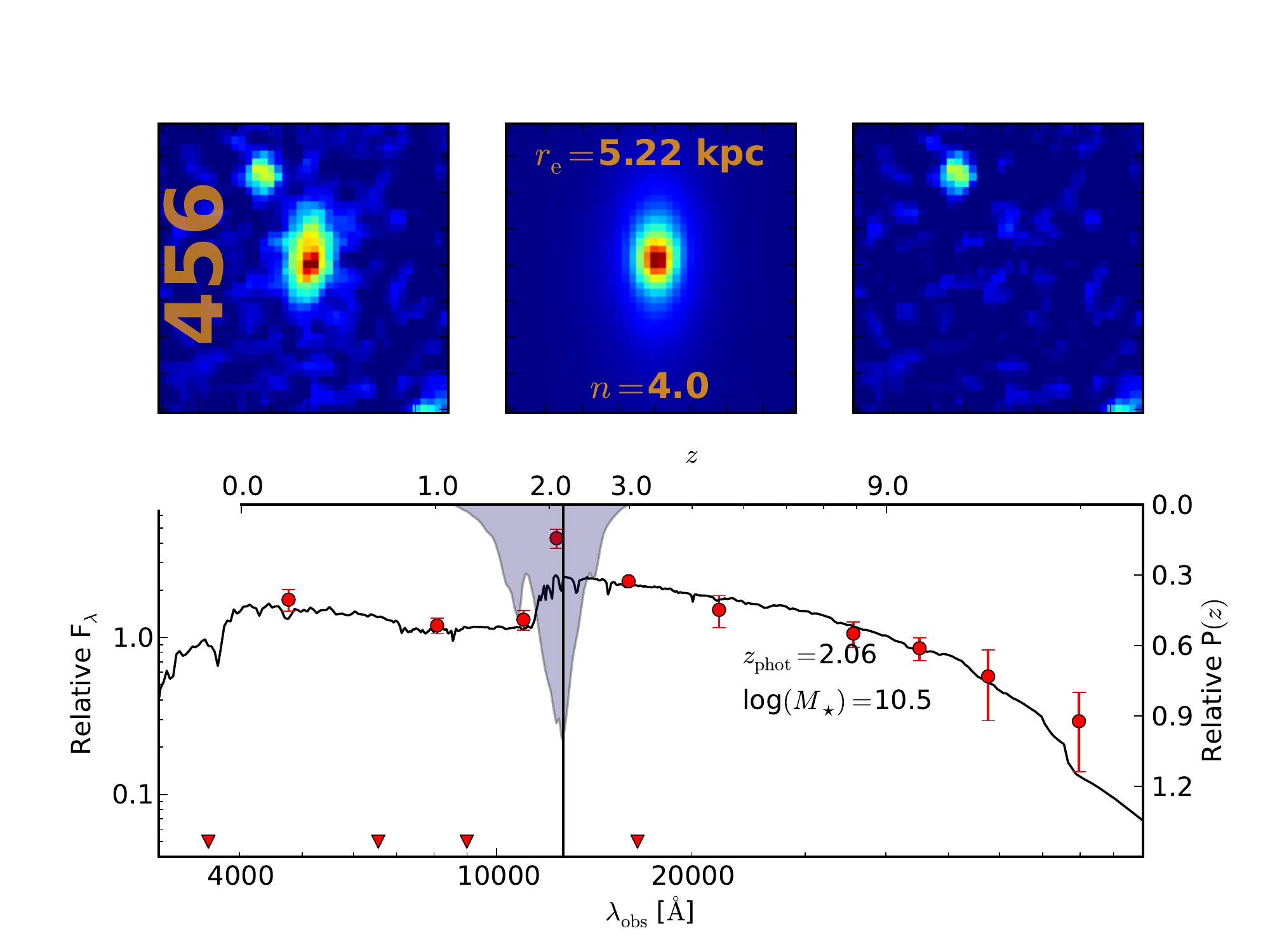}
\includegraphics[scale=0.4]{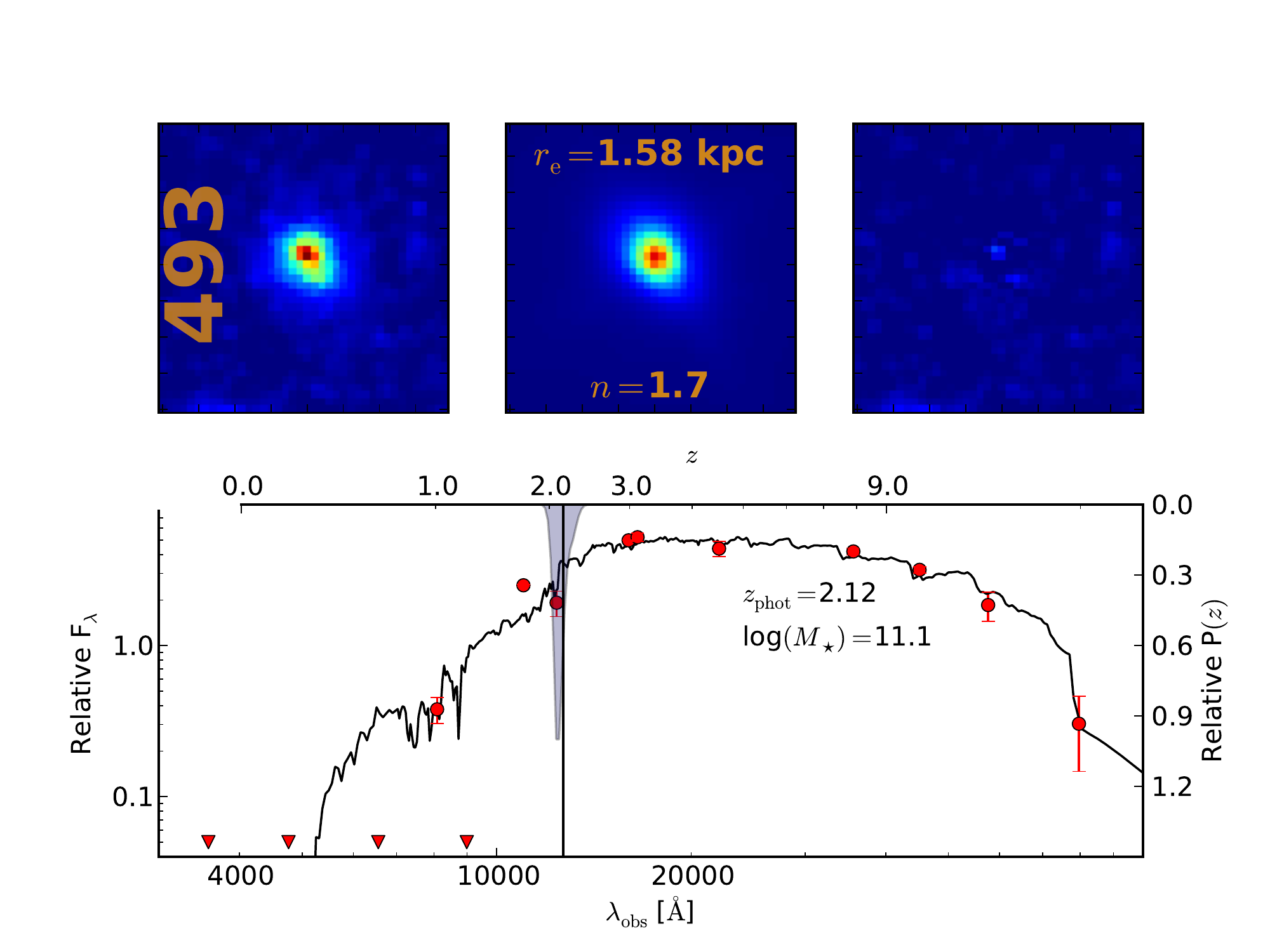}
\includegraphics[scale=0.4]{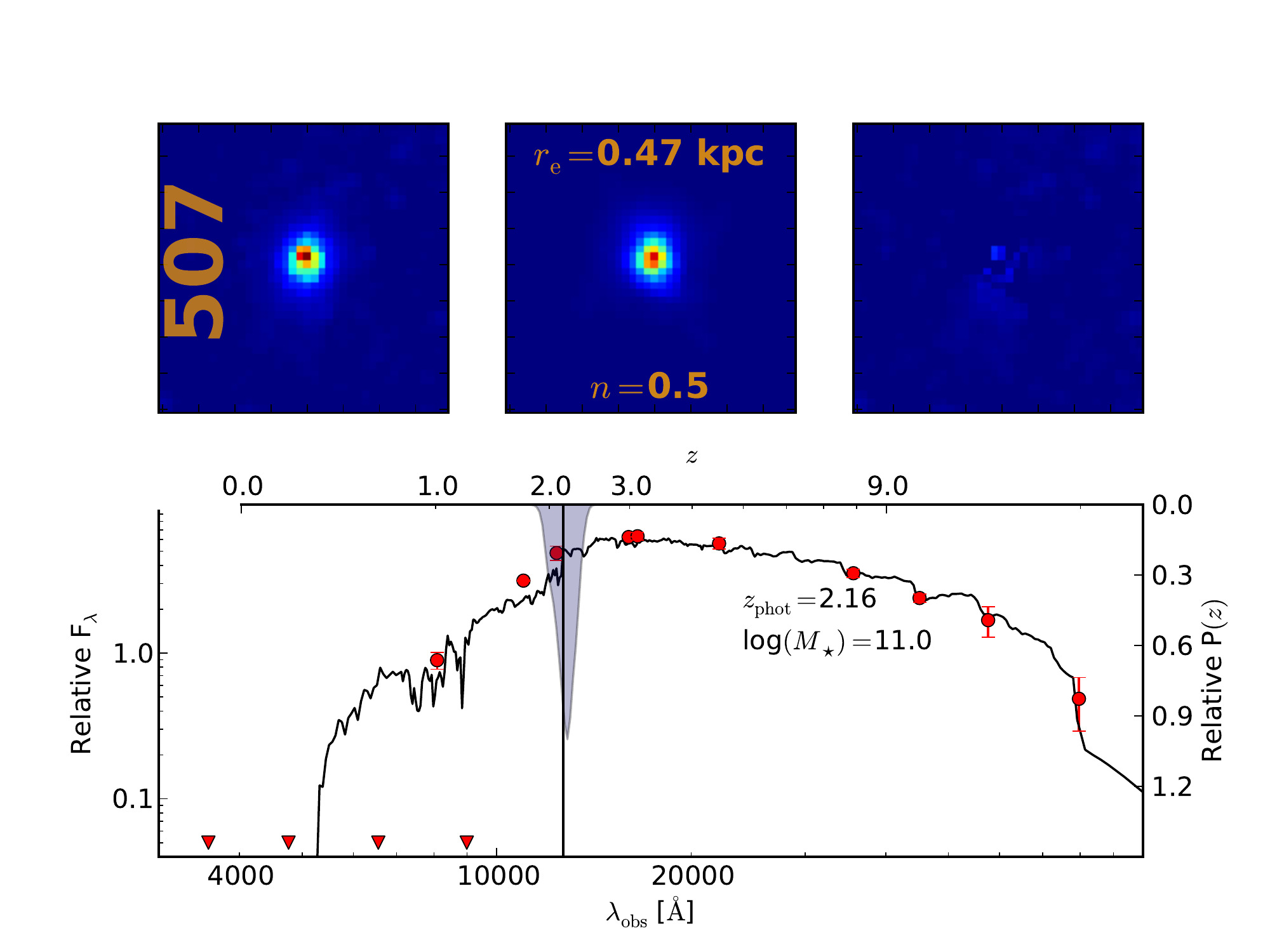}
\includegraphics[scale=0.4]{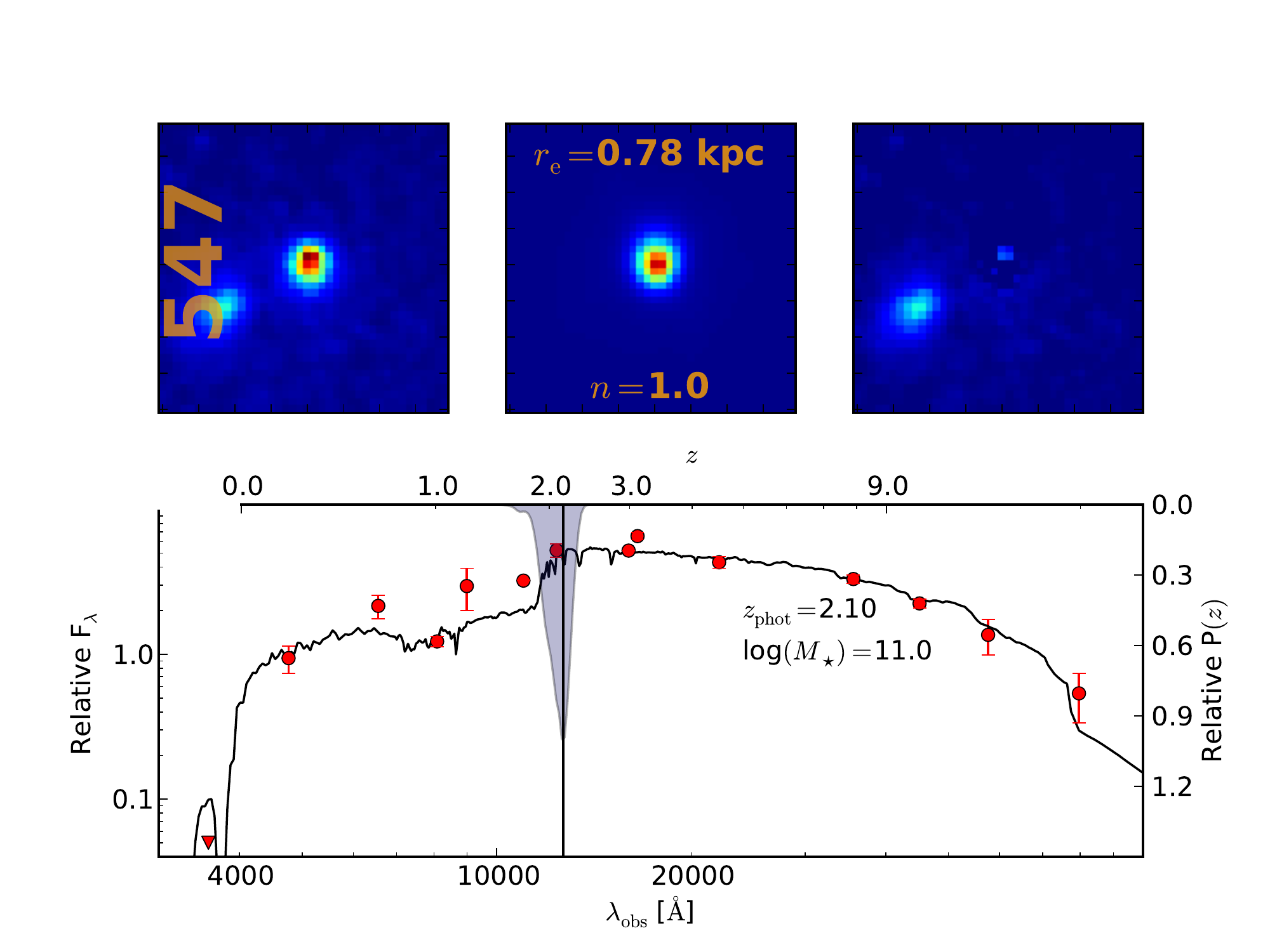}
\caption{Fig. 1 cont.\label{fig:cuts2}}
\end{figure*}

\subsection{Photometric Redshifts\label{sec:EAZY}}

\begin{figure*}
\includegraphics[scale=0.4]{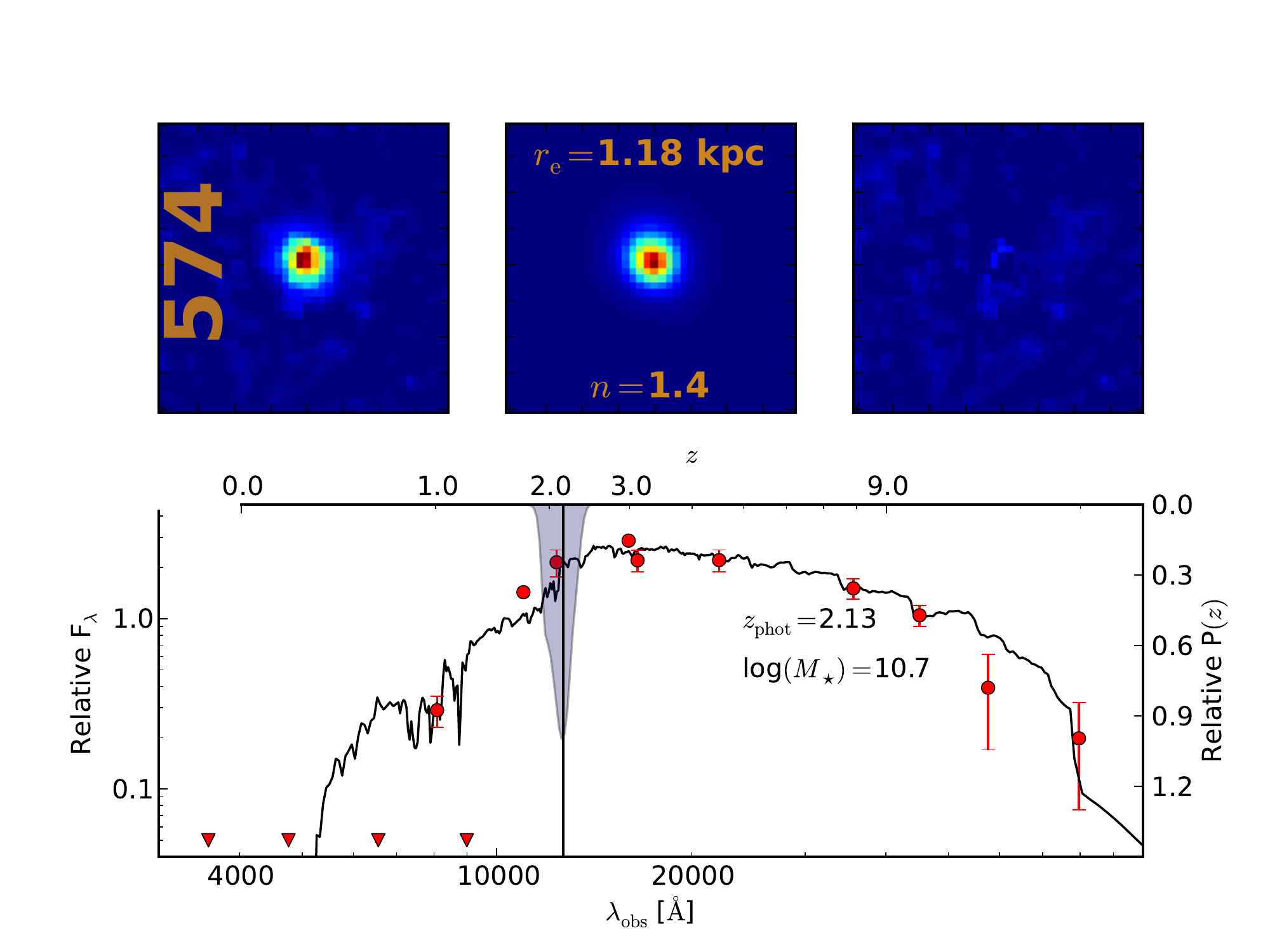}
\includegraphics[scale=0.4]{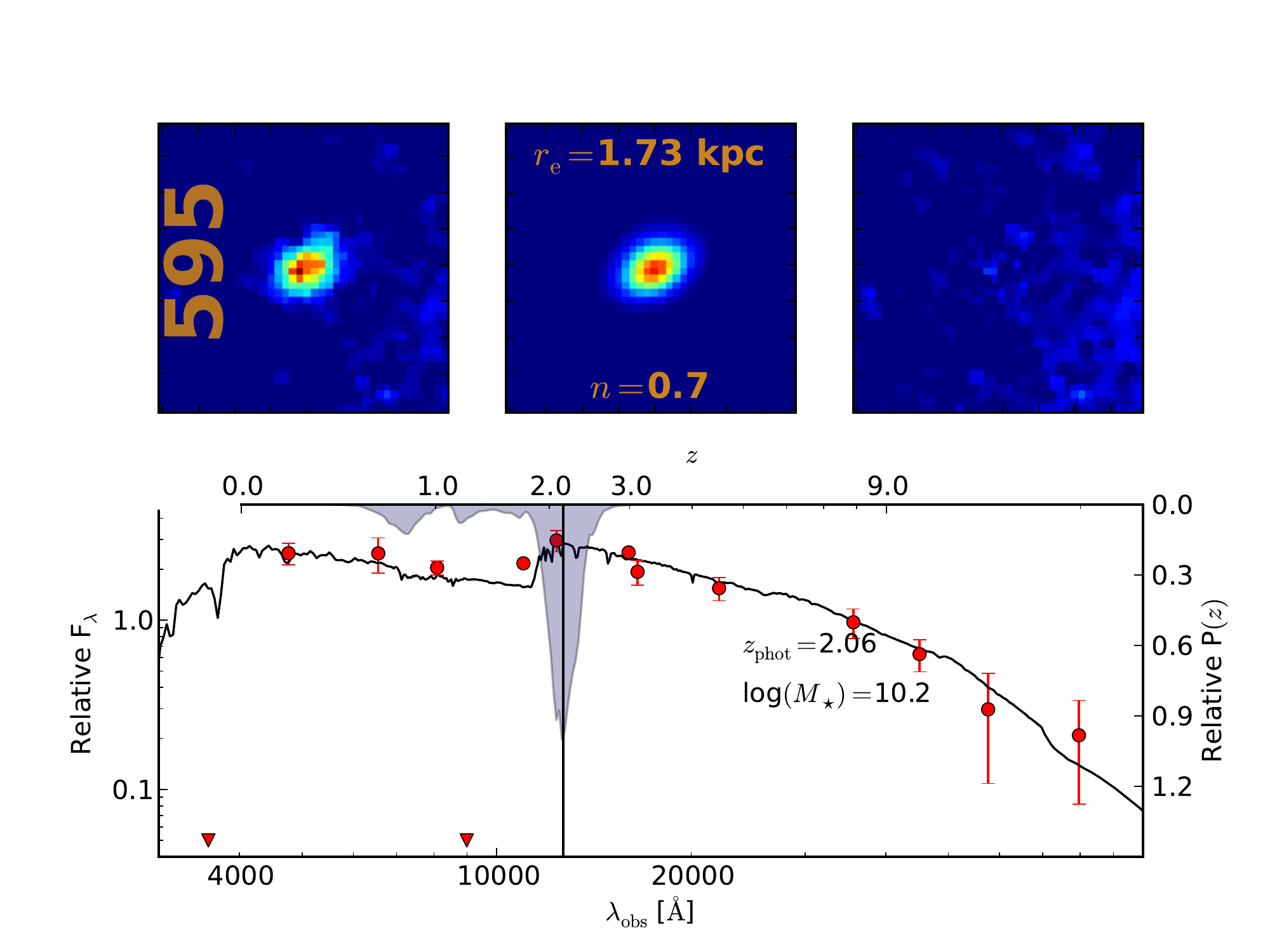}
\includegraphics[scale=0.4]{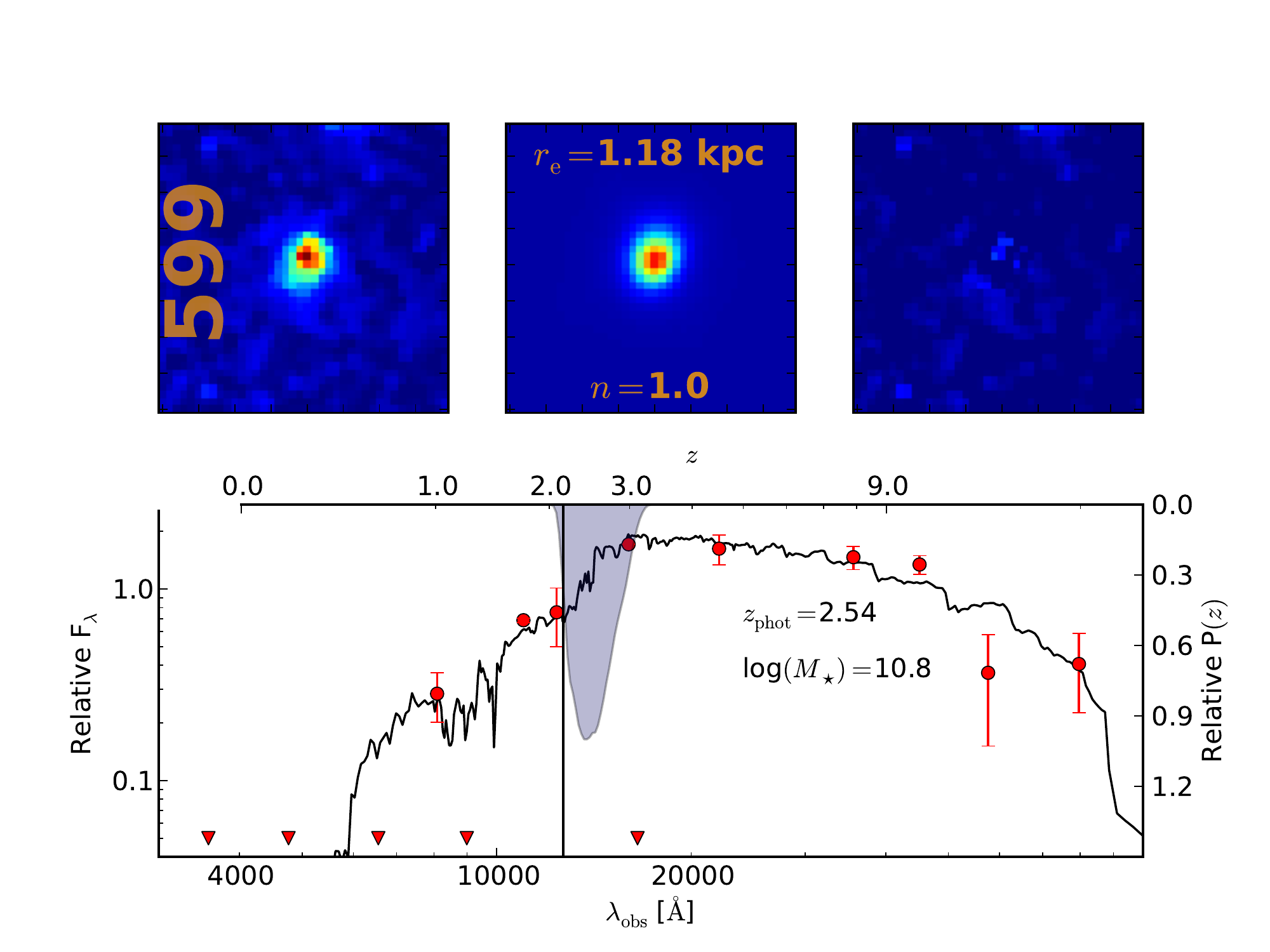}
\caption{Fig. 1 cont.\label{fig:cutslast}}
\end{figure*}

The thirteen filter photometric catalog \citep*{Tanakaetal10} was used
to determine galaxy photometric redshifts.  We used the public code,
EAZY, to fit a set of model templates to each galaxy's photometric
data \citep*{EAZY}.  We required that each galaxy have at least 5
colors measured for the photometric redshift fit.  The set of SED
templates we used included both galaxy spectral energy distributions
and a narrow emission line spectrum.  EAZY uses all linear
combinations of the input templates to find the best photometric
redshift fit.  For each fit, EAZY produces the full redshift
probability distribution (see Figs. \ref{fig:cuts1} -
\ref{fig:cutslast}).  We are most interested in the galaxies detected
in the relatively small ($\sim 5\square\arcmin$), but deep, NICMOS
$H_{160}$-band area.  Therefore, we have only included sources
detected in the $H_{\rm 160}$ NICMOS images.  For the target redshift
of $z = 2.2$, the primary strong spectral feature covered by the
photometric data is the 4000\AA\ break.  We note that even with 13
bands of imaging, photo-$z$s are not sufficiently precise to determine
whether a galaxy is inside the cluster or not.  There are 12
spectroscopically confirmed (emission line) protocluster members
within the NICMOS mosaic.  Of these, four $H\alpha$ and one Ly$\alpha$
emitters have well-determined photometric redshifts (the remaining
members are generally too faint to have detections in enough bands).
Four of the five photo-$z$s are around $z\sim2.1$, ranging from $1.8$
to $2.1$.  There is one clear outlier, the $H\alpha$ emitter with
$z_{\rm phot} = 0.33$.

% FROM MARISKA'S FAQ
% This is caused by the fact that EAZY (or any other photometric
% redshift code) and FAST use different template sets and thus produce
% different PDFs of z. While developing FAST we have explored many
% different methods to derive confidence intervals when EAZY output is
% used. Unfortunately there is no perfect solution, but the one that we
% choose is close and easy to implement in FAST (without writing
% wrappers around EAZY). However, for some galaxies the best-fit
% solution does not fall within the confidence intervals.

\subsection{Stellar Population Modeling\label{sec:FAST}}

Using the calculated best-fit photometric redshifts, we used FAST
\citep*{FAST} to fit stellar population model templates to the
rest-frame photometry.  These templates consist of a grid of models
drawn from the Maraston (2005)\nocite{Maraston05} set.  We chose to
use the Salpeter IMF, exponentially declining star-formation histories
with $\tau$ varying between $10^{7}$ and $10^{10}$ years and $A_{V}$
between $0$ and $3$ magnitudes.  FAST calculates the best-fitting
model template among the grid and thereby derives a
luminosity-weighted mean stellar age, stellar mass, star-formation
rate and extinction for each galaxy. FAST also outputs the 1$\sigma$
error estimates for each of these fit parameters. We show the derived
masses and their errors for the protocluster galaxies in
Table~\ref{tab:data}. We note that the star-formation rates from SED
fitting are equivalent to a dust-corrected rest-frame UV SFR and that
none of our quiescent protocluster galaxies have significant
detections in the MIPS 24~$\mu{\rm m}$ image (20693, PI:
Stanford). Since we are only concerned with differentiating the
quiescent and star-forming galaxies, our results are sensitive only to
catastrophic errors in these SFR determinations.

\subsection{NICMOS Galaxy Sizes and Morphologies\label{sec:morf}}

NICMOS camera 3 provides good angular resolution over its
field-of-view (PSF FWHM $\approx 0\farcs27$).  To exploit this
resolution we have used the GALFIT code \citep*{GALFIT} to fit
analytic S{\'e}rsic surface-brightness profiles \citep*{Sersic} to all
the $H_{160} \leq 26.5$ sources in our $H_{160}$-band mosaic.  We have
used our own error map as input to GALFIT for properly weighting the
image pixels and have masked all neighboring objects.  A model
point-spread function was created for each of these galaxies
individually by generating a TinyTim simulated PSF \citep*{TINYTIM} at
the galaxies' positions in each exposure and then drizzling these PSFs
together in exactly the same fashion as for the data themselves (see
Zirm et al.\ 2007\nocite{Zirmetal07}).  We then executed several
different runs of GALFIT.  We ran fits holding the S{\'e}rsic index
constant at $n=1$ and $4$, using a single model PSF for all galaxies,
using a stellar PSF instead of the model(s) and holding the sky value
fixed at zero.  For all fits we restricted the S{\'e}rsic index, $n$,
to be between 0.5 and 5.  The range of output fit values for all these
different runs gives us an idea of the variance of the derived
parameters due to model assumptions. We choose the best fit from these
runs by applying the F-test to the resultant $\chi^2_{\nu}$ values.

The primary source of systematic offsets in galaxy profile and size
fitting is the estimation of the local sky value. If the sky is
underestimated the galaxy size can be overestimated, particularly for
small faint galaxies. Therefore, we compare our fits where the sky
level is a free parameter with those where we explicitly fix the sky
to zero. Many of the ``zero-sky'' fits fail to converge, for those
that do converge and have comparable chi-squared values to the
corresponding free fits, we can compare the output $r_e$
determinations.  It does not appear to be the case that equally good
fits are obtained with and without fitting the sky. In those five
cases where an F-test shows the zero-sky fit to be better, the sizes
agree within the errors. Furthermore, none of these where the zero-sky
result is comparably good are for any or our protocluster galaxies. We
note that the GALFIT sky values while non-zero are consistently
several orders-of-magnitude smaller than the values corresponding to
galaxy pixels. This sensitivity of the fit parameters to even slight
variations in the sky highlights the importance of fitting the local
sky along with the galaxy parameters (even in sky-subtracted data).
% The cutouts are sufficiently large to allow a good determination of
% the sky value in pixels without appreciable galaxy light.

We also note that additional scatter to the derived $r_e$ values
introduced by using a stellar rather than model PSF is about $10-20\%$
and therefore comparable to the scatter on the single-fit measurements
themselves.

\subsection{Sample Selections\label{sec:res:selection}}

We use the surface-brightness profile and stellar population fits
along with the photometric redshifts to define three sub-samples of
the NICMOS-detected galaxies.  We detail these in order of increasing
restriction. The initial sample is defined by a single $H_{160}$-band
limit of $26.5$ (AB) and consists of 711 galaxies.

\subsubsection{GALFIT Sample}

We ran GALFIT on the full $H_{160}$-limited galaxy sample. Using the
distribution of GALFIT $\chi^2_{\nu}$ values for the fits we identify
galaxies with good quality-of-fit ($\chi^2_{\nu}<2$; 577 of the input
711).  We then calculate the circularized $r_e$ ($= \sqrt{ab}$) for
these well-fit galaxies. We use this GALFIT sample in our analysis of
the dependence of the Sersic index, $n$, on radial position within the
cluster (see left panel of Fig.~\ref{fig:histos}). We note that we
have re-normalized the input sigma (error) maps such that the best
fits have $\chi^2_{\nu} \sim 1$. 

\subsubsection{Stellar Populations Sample}

To select galaxies with good constraints on both the stellar mass and
star-formation rate we have selected another sub-sample for the
initial 711 galaxies.  For targets with both good photo-z fits and
narrow redshift probability distributions (ODDS $>0.90$, meaning 90\%
of the probability distribution is contained within the $\Delta z =
0.2$ around the peak value) the sample comprises 190 galaxies out of
the 711. The photometry for these targets were fit using FAST.  From
this set we have identified 112 with reliably derived parameters
($\chi^2_{\nu} < 3.0$) based on the SED fits (for example, see
Figs.~\ref{fig:cuts1}-\ref{fig:cutslast}).

\subsubsection{Probable Protocluster Galaxies\label{sec:selection}}

Finally, we have identified a sample of candidate protocluster members
using photometric redshifts.  In lieu of spectroscopic redshifts,
which are difficult to obtain for $z \sim 2$ red, quiescent galaxies,
this selection should reject most of the interlopers.  We consider
only galaxies with a robust photometric redshift (as above) that have
probability $P(z) > 20$\% at the protocluster redshift ($z=2.156$).
We further require that these galaxies also are members of the
``Stellar Populations'' sample.  These selections result in a sample
of 11 galaxies, nine of which are quiescent (log$_{10}$ sSFR $< -11$
yr$^{-1}$).  We show the galaxy cutouts, the best-fit Sersic model,
the model subtraction residuals, the broad-band SED fit and
photometric redshift probability distribution for these galaxies in
Figures~\ref{fig:cuts1}-\ref{fig:cutslast}.

\subsection{Stellar Mass Density\label{sec:res:density}}

\begin{figure*}
\plotone{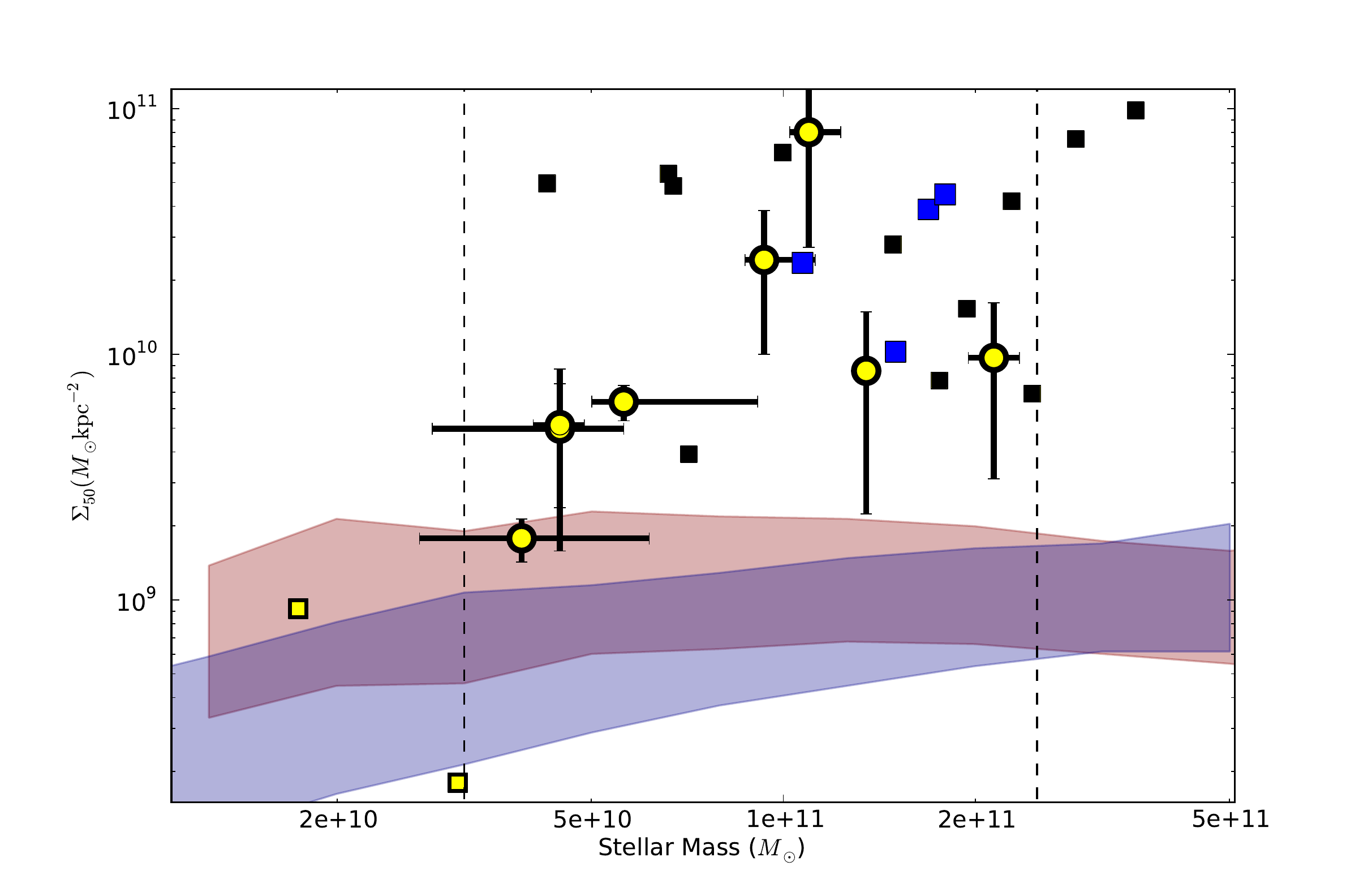}
\caption{Surface stellar mass density ($\Sigma$) vs. total stellar
  mass for individual galaxies.  The blue and black squares are from
  the FIREWORKS survey \citep*{Toftetal09} and other literature
  \citep*{vanDokkumetal08, Cassataetal10, Mancinietal10,
    Saraccoetal09} respectively.  The large blue squares are the
  $K>21.5$ FIREWORKS galaxies included in the KS test. The yellow
  circles with error bars are the 8 likely quiescent protocluster
  members. The two yellow squares are the star-forming protocluster
  galaxies (the third falls at $5 \times 10^{9} M_{\odot}$). The
  shaded regions are the local relations for early-type (light red)
  and late-type galaxies (light blue). Note that most of the
  protocluster members have lower densities than their field
  counterparts.\label{fig:density}}
\end{figure*}

The surface (volume) mass density in individual galaxies is a
fundamental property which seems to correlate directly with the
absence of star formation \citep*[e.g.,][]{Kauffmannetal03,
  Franxetal08}.  To measure this quantity requires accurate total mass
estimates based on either stellar velocity dispersion or, much more
commonly, the stellar mass from SED fits to the broad-band photometry.
Along with the resolved surface-brightness profile to determine the
galaxy size we can calculate the mass density.  We must make the
assumption that light traces mass and that there are no strong
gradients in the stellar mass-to-light ratio, i.e., we measure $M/L$
for the integrated galaxy light and assume that value applies to the
resolved profile in a single broad-band image.

For our galaxies, we calculate the average surface mass density
($\Sigma_{\rm 50}$ in $M_{\odot}$ ${\rm kpc}^{-2}$) within the
(circularized) effective radius ($r_e$) as follows:
\begin{equation}
\Sigma_{\rm 50} = \frac{M_{\star}/2}{\pi r_e^{2}}
\end{equation}

% We have compiled examples from the literature where both the size
% and stellar mass are measured for (quiescent) galaxies.
We present our measurements in Table~\ref{tab:data}.  
% The density-stellar mass diagram is shown in
% Figure~\ref{fig:density}.

\section{Results\label{sec:res}}

\subsection{Distribution of Internal Surface Mass
  Densities\label{sec:density}}

We have used the combination of photometric redshifts, stellar
population modeling and surface-brightness profile fits to calculate
internal surface mass densities for our protocluster sample.
% This galaxy sample includes both field and protocluster galaxies.
We have also added data from the literature and from FIREWORKS to
construct a well-populated density versus stellar mass diagram in
Figure \ref{fig:density}.  If the published data had a measured
star-formation rate in addition to the stellar mass we have restricted
the points plotted to those with low sSFRs (quiescent; ${\rm log_{10}}$
  ${\rm sSFR} < -11$ ${\rm yr^{-1}}$).  In cases where the star-formation
rate was not quoted explicitly, we only plot those galaxies which are
described as ``quiescent'' by the authors.  This distribution for both
our protocluster candidates (yellow circles and squares for quiescent
and star-forming) and the field sources from the literature (black
squares) and FIREWORKS (blue squares) is shown in
Figure~\ref{fig:density}.  For the comparison field sample we have
restricted to FIREWORKS galaxies brighter than $K=21.5$ and with
photo-$z$ between 1.9 and 2.6.  We have made the
same redshift cut for the literature sample. This redshift range
approximately corresponds to a 1 Gyr epoch centered on the
protocluster redshift.

\begin{figure}
\plotone{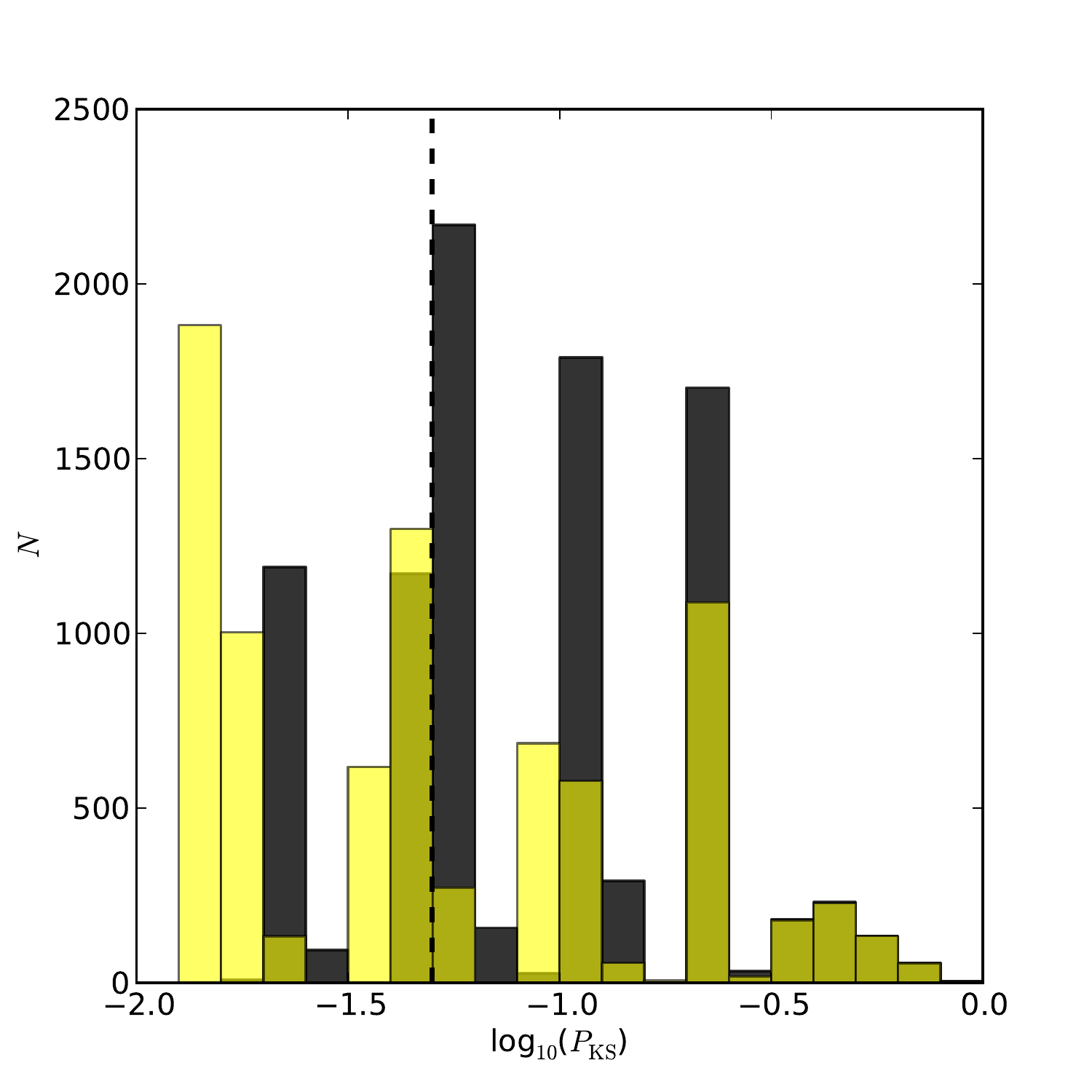}
\caption{The KS probability distributions for the protocluster versus
  field galaxy comparison. The black histogram are derived from 10000
  runs with the protocluster densities perturbed at random within the
  Gaussian errors. The yellow histogram is the same but excluding the
  most dense protocluster galaxy. The vertical dashed line marks
  $P_{\rm KS}=0.05$. About 55\% of the realizations fall below the 5\%
  probability.\label{fig:KS}}
\end{figure}

The mean density of the protocluster sample (${\rm log}$
$<\Sigma_{50}> = 9.9$) is 0.5 dex lower than that for the field
sample.  For the stellar mass range of our protocluster sample,
$10^{10.5}M_{\odot} < M_{\star} < 10^{11.4}M_{\odot}$, where the mass
distributions are similar, we can calculate the distribution of the
surface mass densities irrespective of total stellar mass.  We have
also used the KS test to calculate the probability that the
(quiescent) protocluster and field densities are drawn from the same
parent distribution.  A fiducial value of $P_{\rm KS} < 5\%$ may be
considered sufficient to reject the null hypothesis that they are from
the same parent.  In this case $P_{\rm KS} \sim 5\%$ and is therefore
a relatively strong constraint. We have perturbed our measured
densities and re-calculated $P_{\rm KS}$ for 10000 trials. We show the
distribution of $P_{\rm KS}$ in Figure~\ref{fig:KS}. For the black
(yellow) histogram 30\% (60\%) of the realizations fall below $P_{\rm
  KS} = 5\%$.
% Note that even in the case where we reject the most dense
% protocluster galaxy (yellow histogram) the $P_{\rm KS}$.
Both histograms have tails to higher probabilities.

\subsection{Radial Dependencies\label{sec:radial}}

\begin{figure}
\plottwo{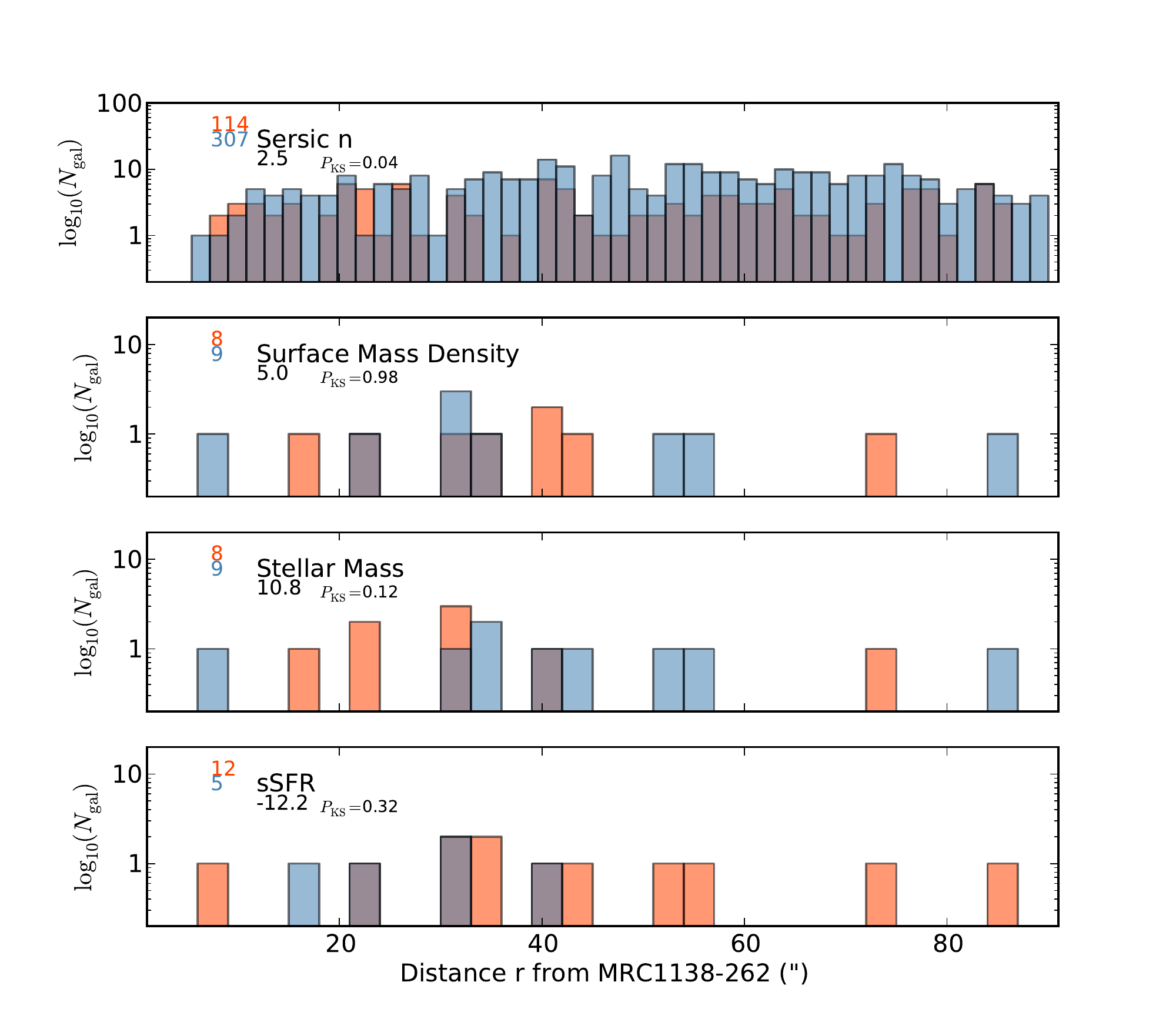}{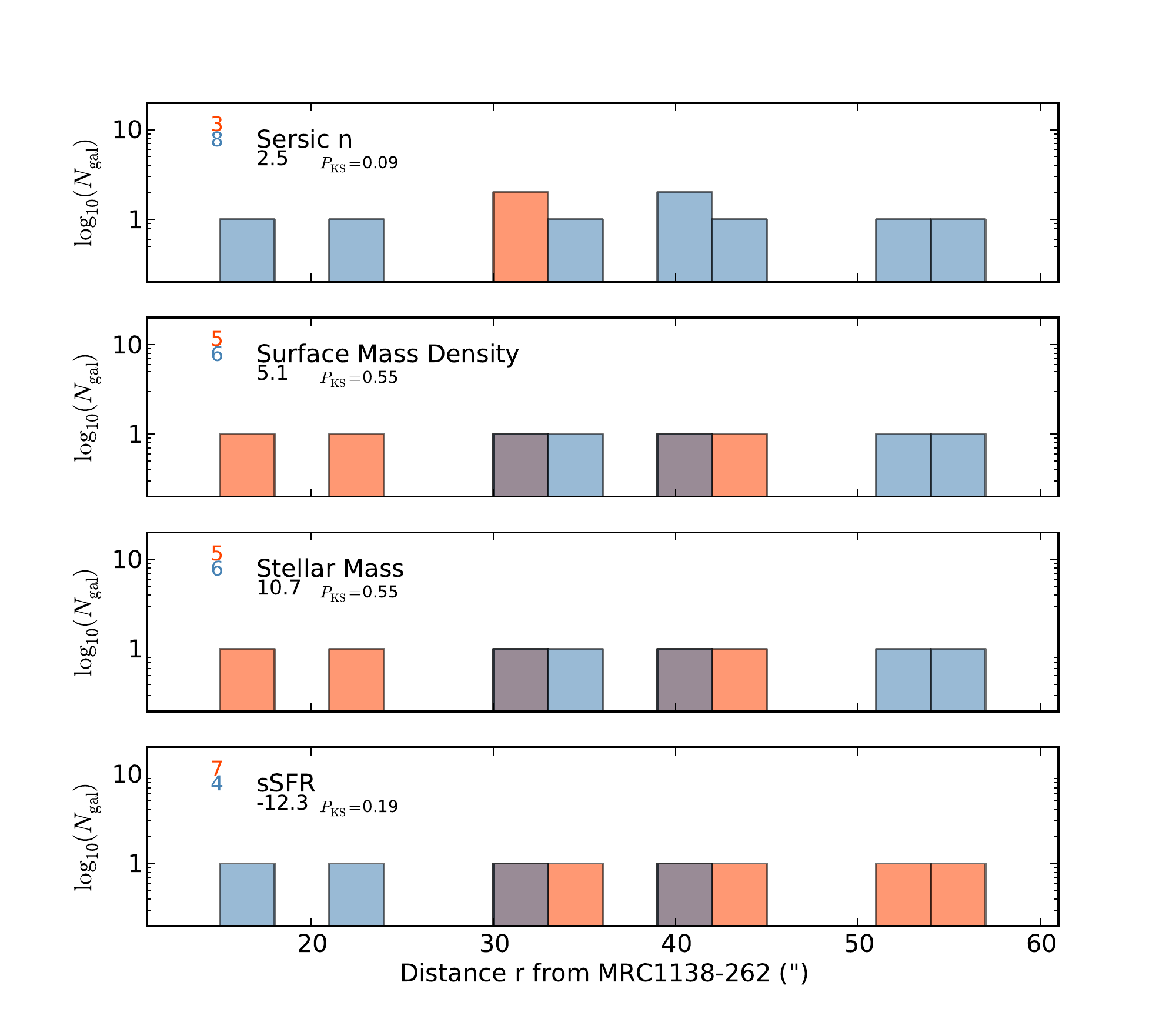}
\caption{Radial (measured from the radio galaxy) distributions of
  galaxies as a function of four derived physical parameters.  Top to
  bottom: S{\'e}rsic $n$ value, surface mass density ($\Sigma$),
  stellar mass ($M_{\star}$) and specific star-formation rate.  For
  each parameter we have made a single cut of the total sample into
  two bins and plotted those two histograms separately for those above
  (orange) and below (blue) this cut.  The value of the cut is shown
  below the parameter name and the number of galaxies in each bin to
  the left of the name.  Finally, the KS probability, that the two
  histograms are drawn from the same parent distribution is also
  shown.\label{fig:histos}}
\end{figure}

In order to assess possible radial gradients in the galaxy properties
within the cluster, with respect to the radio galaxy, we have
constructed the histograms shown in Figure~\ref{fig:histos}.  For each
of these four physical galaxy parameters: S{\'e}rsic $n$ value,
surface mass density ($\Sigma$), stellar mass ($M_{\star}$) and
specific SFR, we have made a single cut in the galaxy sample and
plotted two radial histograms for above (orange) and below (blue) the
chosen cut value.  The left panel of Fig.~\ref{fig:histos} shows the
histograms for the full galaxy sample appropriate to each parameter,
i.e., the GALFIT sample for the Sersic $n$ value.  For the rest, the
galaxies must also be in the stellar population sample. The right
panel shows the histograms for our protocluster galaxy sample (both
quiescent and star-forming). We have presented these two analyses
because while the presence of field galaxies in the larger samples
will dilute any result, the statistics are poor for the quiescent
protocluster galaxy sample.  Furthermore, this field exhibits a factor
of six surface overdensity of red galaxies \citep*{Zirmetal08}, so the
large sample may not strongly dilute trends.

For each pair of histograms we have run the two-sample KS test to
determine whether the distributions are consistent with one another.
However, for the full sample (left) the Sersic $n$ distributions are
more dissimilar than for the other parameters with a low $P_{\rm KS} =
5\%$. This low probability seems to be the result of a relatively flat
distribution of the $n < 2.5$ galaxies with radius contrasting with
the structure in the radial distribution of the higher $n$
galaxies. This hint may imply a scenario in which the galaxies are
deeper within the gravitational potential.  We discuss this point
further below.  In the future, with spectroscopic redshifts for red
galaxies, it will be possible to repeat this test with better
interloper rejection to see if the discrepancy between histograms is
significant for bona fide protocluster galaxies. 

\section{Discussion\label{sec:discussion}}

We have presented the combined analysis of 13 band photometry and high
spatial-resolution NIR imaging in the field of a known galaxy
protocluster at $z=2.16$. 
% Studies of the complete protocluster galaxy population, i.e.,
% particularly the quiescent red galaxies, are somewhat hampered by
% the lack of spectroscopic redshifts.
In cases like this where there is a known, confirmed overdensity and
strong statistical evidence for a dominant contribution from
protocluster members we can make progress despite the lack of
spectrocopic redshifts.  We have identified a robust sample of likely
protocluster galaxies. Our conclusions are tentative and tempered by
the following caveats. We do not have spectroscopic redshifts for the
quiescent cluster galaxies. While we have done our best to isolate the
most probable cluster members, they may still be field galaxies. The
cluster sample is also relatively small and the results are therefore
more suggestive rather than statistically robust. Finally, due to the
limited areal coverage of the protocluster (see
Fig.~\ref{fig:spatial}), the very massive galaxies may be
underrepresented in the cluster sample. These may also be more dense.

Possible stellar population model offsets have been minimized by
converting the literature points to the same models and IMF we used to
fit the cluster galaxies.
% With such a sample cultivated from our extensive data in the field
% of MRC 1138-262, we can begin to address differences in galaxy
% structure, size and morphology inside and outside this known
% overdensity.  Here we discuss the results of this investigation.

\subsection{Evolution of Galaxy Structure and Stellar Mass Surface
  Density\label{sec:densities}}

From the initial discovery of SEEDs, in general field surveys, the
primary question has been what evolutionary processes affect the SEEDs
between $z \sim 2-3$ and $z=0$ that bring them in line with the local
mass-size relation.  If the dominant process is galaxy merging, then
we might expect that most of the full galaxy mergers, as opposed to
tidal interactions and harassment, may have already happened. While if
the primary determinant of galaxy density is the formation redshift,
the young Universe being denser, we might expect that cluster galaxies
will be denser than their field counterparts having formed earlier.

In our data we see some indication for a difference between the
profile shapes and density distributions for protocluster versus field
galaxies (Fig.~\ref{fig:density}).  For our sample of likely quiescent
protocluster galaxies their stellar densities are lower and perhaps
even the Sersic index is higher than for similarly selected field
galaxies.  From other studies it appears that the majority of field
SEEDs have higher axial ratios (flattened) with $n \sim 2$
\citep*{vanderWeletal11}.

\subsection{Cosmic Merger Clocks\label{sec:clocks}}

Several previous studies have shown evidence that protocluster
galaxies tend to be more massive and contain older stars than their
field counterparts at the same redshift
\citep*[e.g.,][]{Steideletal05, Tanakaetal10}. This advanced evolution
in the cluster environment may also extend to the internal structure
and dynamics of the galaxies. At lower redshift, we observe a
morphology-density relation, and we may be seeing the beginnings of
that relation in the MRC~1138 protocluster. Furthermore, the lower
densities of the protocluster galaxies suggests that the necessary
merging has also taken place at a quicker pace than in the field. If
we assume then that all galaxies begin as dense SEEDs at high
redshift, we can use the observed densities as a measure of the
``merger age'' of the remnants. A similar idea was put forth by
Hopkins and Hernquist (2010)\nocite{HopkinsHernquist10} to derive the
global star-formation history by using the mass profiles of
galaxies. They propose that the dense cores of galaxies form early in
starbursts and the outer parts form in a more quiescent mode, perhaps
in disks. Based on the results from this paper we suggest that the
resolved mass profiles can be used further to constrain the redshift
of formation and the subsequent merger history. The ratio of
high-density stellar mass to low-density stellar mass may tell us
something about the merger age of a galaxy while the absolute
density of the highest density stellar components may tell us about
the formation redshift. Studies at low-redshift have already found
some correlation between galaxy density and the mean stellar age
\citep*[e.g.,][]{vanderWeletal09}. With more precise stellar ages for
$z \sim 2$ galaxies now becoming available we can extend this analysis
to high-redshift when the fractional age differences between galaxies
are larger.

\acknowledgments 

A. Zirm and S. Toft gratefully acknowledge support from the Lundbeck
Foundation. The authors thank the anonymous referee for their helpful
comments. This work is supported by World Premier International
Research Center Initiative (WPI Initiative), MEXT, Japan and also in
part by Grant-In-Aid for Young Scientists No. 23740144.

\bibliographystyle{apjv2}
%\bibliography{journals,1138dens,BIG,1138CMR,1138CMR_other,1138CMR_other2}

\newpage

\input{table_test.tex}

\end{document}

%% file: table_test.tex
\begin{deluxetable}{rrrrrrrrrr}
    \tabletypesize{\scriptsize}
    \tablecolumns{10} 
    \tablewidth{0pc} 
    \tablecaption{Protocluster Candidates \label{tab:data}} 
    \tablehead{ 
    \colhead{Object} & \colhead{Photometric} & \colhead{Odds} & \colhead{$H_{160}$} & \colhead{Line} & \colhead{Stellar} & \colhead{Specific} & \colhead{Effective} & \colhead{Effective} & \colhead{Mass} \\
\colhead{Object} & \colhead{Redshift} & \colhead{} & \colhead{(AB)} & \colhead{} & \colhead{Mass} & \colhead{SFR} & \colhead{Radius} & \colhead{Radius} & \colhead{Density} \\
\colhead{} & \colhead{} & \colhead{} & \colhead{} & \colhead{} & \colhead{(log($M_{\odot}$))} & \colhead{(log(yr$^{-1}$))} & \colhead{($\arcsec$)} & \colhead{(kpc)} & \colhead{($10^{9}$ $M_{\odot} {\rm kpc}^{-2}$)}
}
\startdata
\cutinhead{Quiescent Protocluster Galaxies}
312 & 2.24 & 1.00 & 21.98 $\pm$ 0.01 & 0 & $11.33^{+0.04}_{-0.04}$ & -99.00 & 0.23 $\pm$ 0.09 & 1.89 & $9.67 \pm 6.5$ \\ 
394 & 2.23 & 1.00 & 24.07 $\pm$ 0.01 & 0 & $10.65^{+0.20}_{-0.20}$ & -11.24 & 0.14 $\pm$ 0.07 & 1.20 & $4.98 \pm 2.6$ \\ 
399 & 2.49 & 0.98 & 24.20 $\pm$ 0.01 & 0 & $10.59^{+0.20}_{-0.16}$ & -11.24 & 0.23 $\pm$ 0.09 & 1.91 & $1.78 \pm 0.4$ \\ 
493 & 2.12 & 1.00 & 22.73 $\pm$ 0.01 & 0 & $11.13^{+0.03}_{-0.04}$ & -99.00 & 0.19 $\pm$ 0.07 & 1.58 & $8.56 \pm 6.2$ \\ 
507 & 2.16 & 1.00 & 22.48 $\pm$ 0.01 & 0 & $11.04^{+0.05}_{-0.03}$ & -99.00 & 0.06 $\pm$ 0.06 & 0.47 & $80.14 \pm 52.9$ \\ 
547 & 2.10 & 1.00 & 22.68 $\pm$ 0.01 & 0 & $10.97^{+0.08}_{-0.03}$ & -11.72 & 0.09 $\pm$ 0.08 & 0.78 & $24.21 \pm 14.2$ \\ 
574 & 2.13 & 1.00 & 23.33 $\pm$ 0.01 & 0 & $10.65^{+0.04}_{-0.04}$ & -99.00 & 0.14 $\pm$ 0.09 & 1.17 & $5.15 \pm 3.6$ \\ 
599 & 2.54 & 1.00 & 23.89 $\pm$ 0.01 & 0 & $10.75^{+0.21}_{-0.05}$ & -12.20 & 0.15 $\pm$ 0.08 & 1.22 & $6.41 \pm 1.0$ \\ 
%660 & 2.03 & 1.00 & 24.16 $\pm$ 0.02 & 0 & $10.91^{+0.03}_{-0.05}$ & -99.00 & 0.38 $\pm$ 0.11 & 3.17 & $1.27 \pm 0.9$ \\ 
\cutinhead{Star-forming Protocluster Galaxies}
347 & 2.08 & 0.92 & 23.69 $\pm$ 0.01 & 0 & 9.69 & -8.79 & 0.19 $\pm$ 0.05 & 1.57 & 0.31 \\ 
456 & 2.06 & 0.95 & 23.57 $\pm$ 0.07 & 0 & 10.49 & -9.27 & 0.63 $\pm$ 0.06 & 5.19 & 0.18 \\ 
595 & 2.06 & 0.91 & 23.47 $\pm$ 0.01 & 0 & 10.24 & -8.79 & 0.21 $\pm$ 0.07 & 1.72 & 0.92 \\ 
\cutinhead{Line-emitting Candidates\tablenotemark{a}}
700 & 2.16 &  \nodata  & 24.02 $\pm$ 0.02 & 1 &  \nodata  &  \nodata  & 0.23 $\pm$ 0.06 & 1.88 &  \nodata  \\ 
463 & 2.16 &  \nodata  & 24.51 $\pm$ 0.03 & 1 &  \nodata  &  \nodata  & \nodata  &  \nodata  &  \nodata  \\ 
1078 & 2.16 &  \nodata  & 22.94 $\pm$ 0.01 & 1 &  \nodata  &  \nodata  & 0.17 $\pm$ 0.06 & 1.38 &  \nodata  \\ 
1070 & 2.16 &  \nodata  & 23.44 $\pm$ 0.02 & 1 &  \nodata  &  \nodata  & \nodata  &  \nodata  &  \nodata  \\ 
511 & 2.16 &  \nodata  & 24.50 $\pm$ 0.01 & 1 &  \nodata  &  \nodata  & \nodata  &  \nodata  &  \nodata  \\ 
516 & 2.16 &  \nodata  & 24.07 $\pm$ 0.01 & 1 &  \nodata  &  \nodata  & 0.14 $\pm$ 0.08 & 1.15 &  \nodata  \\ 
575 & 2.16 &  \nodata  & 24.62 $\pm$ 0.02 & 1 &  \nodata  &  \nodata  & 0.01 $\pm$ 0.03 & 0.09 &  \nodata  \\ 
988 & 2.16 &  \nodata  & 23.41 $\pm$ 0.01 & 1 &  \nodata  &  \nodata  & 0.32 $\pm$ 0.06 & 2.67 &  \nodata  \\ 
536 & 2.16 &  \nodata  & 23.90 $\pm$ 0.01 & 1 &  \nodata  &  \nodata  & 0.13 $\pm$ 0.08 & 1.11 &  \nodata  \\ 
1069 & 2.16 &  \nodata  & 23.91 $\pm$ 0.01 & 1 &  \nodata  &  \nodata  & 0.00 $\pm$ 0.54 & 0.01 &  \nodata  \\ 
457 & 2.16 &  \nodata  & 23.84 $\pm$ 0.01 & 1 &  \nodata  &  \nodata  & \nodata  &  \nodata  &  \nodata  \\ 
451 & 2.16 &  \nodata  & 23.74 $\pm$ 0.01 & 1 &  \nodata  &  \nodata  & \nodata  &  \nodata  &  \nodata  \\ 
275 & 2.16 &  \nodata  & 23.62 $\pm$ 0.01 & 1 &  \nodata  &  \nodata  & 0.12 $\pm$ 0.06 & 0.96 &  \nodata  \\ 
897 & 2.16 &  \nodata  & 24.09 $\pm$ 0.01 & 1 &  \nodata  &  \nodata  & 0.11 $\pm$ 0.05 & 0.92 &  \nodata  \\ 
300 & 2.16 &  \nodata  & 22.50 $\pm$ 0.01 & 1 &  \nodata  &  \nodata  & \nodata  &  \nodata  &  \nodata  \\ 
311 & 2.16 &  \nodata  & 23.94 $\pm$ 0.01 & 1 &  \nodata  &  \nodata  & 6.77 $\pm$ 4.36 & 56.15 &  \nodata  \\ 
361 & 2.16 &  \nodata  & 24.50 $\pm$ 0.01 & 1 &  \nodata  &  \nodata  & 0.03 $\pm$ 0.04 & 0.27 &  \nodata  \\ 
365 & 2.16 &  \nodata  & 26.16 $\pm$ 0.06 & 1 &  \nodata  &  \nodata  & \nodata  &  \nodata  &  \nodata  \\ 
215 & 2.16 &  \nodata  & 24.09 $\pm$ 0.01 & 1 &  \nodata  &  \nodata  & 0.14 $\pm$ 0.08 & 1.18 &  \nodata  \\ 
448 & 2.16 &  \nodata  & 21.93 $\pm$ 0.02 & 1 &  \nodata  &  \nodata  & \nodata  &  \nodata  &  \nodata  \\ 
435 & 2.16 &  \nodata  & 23.74 $\pm$ 0.01 & 1 &  \nodata  &  \nodata  & \nodata  &  \nodata  &  \nodata  \\ 
431 & 2.16 &  \nodata  & 22.61 $\pm$ 0.01 & 1 &  \nodata  &  \nodata  & \nodata  &  \nodata  &  \nodata  \\ 
\cutinhead{Confirmed Line-emitting Galaxies}
53 & 2.16 &  \nodata  & 24.00 $\pm$ 0.01 & 2 &  \nodata  &  \nodata  & 0.02 $\pm$ 0.04 & 0.19 &  \nodata  \\ 
263 & 2.16 &  \nodata  & 24.01 $\pm$ 0.01 & 2 &  \nodata  &  \nodata  & 0.18 $\pm$ 0.08 & 1.52 &  \nodata  \\ 
270 & 2.16 &  \nodata  & 21.11 $\pm$ 0.01 & 2 &  \nodata  &  \nodata  & \nodata  &  \nodata  &  \nodata  \\ 
450 & 2.16 &  \nodata  & 22.75 $\pm$ 0.01 & 2 &  \nodata  &  \nodata  & \nodata  &  \nodata  &  \nodata  \\ 
945 & 2.16 &  \nodata  & 23.79 $\pm$ 0.01 & 2 &  \nodata  &  \nodata  & 0.03 $\pm$ 0.02 & 0.22 &  \nodata  \\ 
289 & 2.16 &  \nodata  & 22.05 $\pm$ 0.01 & 2 & 11.52 & -11.24 & 0.43 $\pm$ 0.09 & 3.53 & 4.22 \\ 
757 & 2.16 &  \nodata  & \nodata  & 2 & 10.54 & -9.27 & \nodata  &  \nodata  &  \nodata  \\ 
561 & 2.16 &  \nodata  & 21.99 $\pm$ 0.01 & 2 & 11.23 & -12.20 & 0.86 $\pm$ 0.09 & 7.09 & 0.54 \\ 
296 & 2.16 &  \nodata  & 22.27 $\pm$ 0.01 & 2 & 10.50 & -8.18 & 0.10 $\pm$ 0.07 & 0.84 & 7.10 \\ 
648 & 2.16 &  \nodata  & 23.38 $\pm$ 0.01 & 2 &  \nodata  &  \nodata  & 0.14 $\pm$ 0.06 & 1.18 &  \nodata  \\ 
535 & 2.16 &  \nodata  & 23.90 $\pm$ 0.01 & 2 &  \nodata  &  \nodata  & 0.14 $\pm$ 0.05 & 1.19 &  \nodata  \\ 
387 & 2.16 &  \nodata  & 24.70 $\pm$ 0.02 & 2 &  \nodata  &  \nodata  & 0.06 $\pm$ 0.04 & 0.47 &  \nodata  \\ 
\enddata
\tablenotetext{a}{Narrow-band selected objects that are not yet spectroscopically confirmed.}
\end{deluxetable}